\documentclass[reprint,twocolumn,aps,prb,showpacs]{revtex4-1}

\usepackage{epstopdf}
\usepackage{subfigure}

\usepackage{amsmath}
\usepackage{amssymb}
\usepackage{graphicx}
\usepackage{dcolumn}
\usepackage{bm}
\usepackage[colorlinks=true, allcolors=blue]{hyperref}
\usepackage{epstopdf}
\usepackage[utf8]{inputenc}
\usepackage{amssymb}

\begin{document}

\title{Correlation between the strength of low-temperature T-linear resistivity
and $T_{\rm c}$ in overdoped electron-doped cuprate superconductors}

\author{Xingyu Ma$^{\rm a}$, Minghuan Zeng$^{\rm b}$, Huaiming Guo$^{\rm c}$, and
Shiping Feng$^{\rm a}$}
\thanks{Corresponding author. Email: spfeng@bnu.edu.cn}

\affiliation{$^{\rm a}$Department of Physics, Faculty of Arts and Sciences, Beijing
Normal University, Zhuhai 519087, China and\\
School of Physics and Astronomy, Beijing Normal University, Beijing 100875, China}

\affiliation{$^{\rm b}$College of Physics, Chongqing University, Chongqing 401331,
China}

\affiliation{$^{\rm c}$School of Physics, Beihang University, Beijing 100191, China}

\begin{abstract}
The recently observed an intimate link between the nature of the strange metallic
normal-state and superconductivity in the overdoped electron-doped cuprate
superconductors is calling for an explanation. Here the intrinsic correlation
between the strength of the low-temperature linear-in-temperature (T-linear)
resistivity and superconducting transition temperature $T_{\rm c}$ in the overdoped
electron-doped cuprate superconductors is studied within the framework of the
kinetic-energy-driven superconductivity. On the one hand, the main ingredient is
identified into a electron pairing mechanism involving {\it the spin excitation},
and then $T_{\rm c}$ has a dome-like shape doping dependence with the maximal
$T_{\rm c}$ that occurs at around the optimal electron doping. On the other hand,
in the normal-state above $T_{\rm c}$, the low-temperature T-linear resistivity in
the overdoped regime arises from the momentum relaxation due to the electron
umklapp scattering mediated by {\it the same spin excitation}. This {\it same spin
excitation} that governs both the electron umklapp scattering responsible for the
low-temperature T-linear resistivity and electron pairing responsible for
superconductivity naturally generates a correlation between the strength of the
low-temperature T-linear resistivity and $T_{\rm c}$ in the overdoped regime.
\end{abstract}


\maketitle

\section{Introduction}\label{Introduction}

The undoped parent compounds of cuprate superconductors are antiferromagnetic (AF)
Mott insulators \cite{Fujita12}, which result from the particularly strong electron
correlation \cite{Anderson87}. However, the exceptionally strong superconductivity
can be achieved when the AF long-range order (AFLRO) is destroyed by a small
fraction of the electron or hole doping concentration \cite{Bednorz86,Tokura89}. In
particular, although both the electron doping and annealing process in a low-oxygen
environment are required to induce superconductivity in the electron-doped cuprate
superconductors \cite{Armitage10,Adachi13,Adachi17}, the experimental observations
demonstrate unambiguously that in a optimal annealing condition, the
superconducting (SC) phase is extended over a wide electron doping range
\cite{Horio16,Song17,Lin20,Horio20b}. In the recent measurements on the
electron-doped cuprate superconductor Pr$_{1-x}$LaCe$_{x}$CuO$_{4-\delta}$ under
the proper annealing condition \cite{Song17}, it was observed experimentally that
the SC transition temperature $T_{\rm c}$ reaches its maximum at around the optimal
electron doping $\delta\sim 0.15$, while the disappearance of AFLRO at the doping
is coincident with the onset of superconductivity, and hence the deduced AFLRO
phase boundary does not extend into the SC dome. However, the very recent results
\cite{Lin20,Horio20b} observed experimentally from the electron-doped cuprate
superconductor Pr$_{1.3-x}$La$_{0.7}$Ce$_{x}$CuO$_{4-\delta}$ under the protect
annealing condition indicated that the actual electron doping concentration
estimated from the electron Fermi surface (EFS) area is significantly larger than
the Ce doping concentration $x$, and then the new electron-doping-based SC dome is
more extended towards the overdoped side than the Ce-doping-based SC dome obtained
for samples prepared in the proper annealing condition \cite{Song17}. In particular,
although the maximal $T_{\rm c}$ occurs at the optimally doped regime, this
optimally doped regime spans a narrow doped range centered at around the doping
$\delta\sim 0.15$ \cite{Lin20}. Moreover, in the underdoped regime, $T_{\rm c}$
drops rapidly and vanishes at around the doping $\delta\sim 0.10$, while the AF
correlation is found to coexist with superconductivity \cite{Lin20}. Since the
dramatic suppression of the AFLRO has been observed in the underdoped regime of
the electron-doped cuprate superconductors under the protect annealing condition
\cite{Horio16}, one may anticipate a $T_{\rm c}$ recovery for $\delta\leq 0.10$
with further improved condition of the protect annealing \cite{Lin20}. These
experimental results of the phase diagram for the electron-doped cuprate
superconductors show that although some features in the underdoped regime observed
in Ref. \cite{Song17} are significantly different from that in Ref. \cite{Lin20},
the doping range of these phase diagrams \cite{Horio16,Song17,Lin20,Horio20b} bear
a striking similarity to that of the corresponding phase diagram in the hole-doped
counterparts \cite{Drozdov18,Taillefer10}, providing a clue for understanding the
symmetry/asymmetry of the phase diagrams of the electron- and hole-doped cuprate
superconductors. This similarity of the phase diagrams
\cite{Horio16,Song17,Lin20,Horio20b,Drozdov18,Taillefer10} also suggests that the
essential physics, including the SC mechanism and the nature of the normal-state
in {\it the overdoped regime}, is most likely the same for both the electron- and
hole-doped cuprate superconductors.

After intensive investigations over thirty years, a substantial amount of reliable
and reproducible data for the electron-doped cuprate superconductors has been
accumulated through systematic measurements using various techniques
\cite{Taillefer10,Li07,Poniatowski21,Poniatowski21a,Greene20,Fournier98,Jin11,Sarkar17,Sarkar19,Legros19,Yuan22},
which indicate that the normal-state property of these materials is a remarkable
mystery and the most important unsolved problem in condensed matter physics. In
particular, the normal-state property
\cite{Greene20,Fournier98,Jin11,Sarkar17,Sarkar19,Legros19,Yuan22} in the
overdoped regime deviates from the conventional Fermi-liquid behavior
\cite{Schrieffer64,Abrikosov88,Mahan81}. This has led to the normal-state in the
overdoped regime being refereed to as {\it the strange metallic normal-state}
\cite{Keimer15,Varma20,Phillips22}. One of the key manifestations of this deviation
is the low-temperature linear-in-temperature (T-linear) resistivity, which persists
down to millikelvin temperatures and extrapolates to zero resistivity at zero
temperature
\cite{Greene20,Fournier98,Jin11,Sarkar17,Sarkar19,Legros19,Yuan22}.
Recently, this low-temperature T-linear resistivity in the overdoped electron-doped
cuprate superconductors has been confirmed experimentally all the way up to the
edge of the SC dome \cite{Yuan22}. More importantly, these experimental observations
also establish definitively that in the overdoped regime, the low-temperature
T-linear resistivity is tied to $T_{\rm c}$, where the strength of the
low-temperature T-linear resistivity (then the T-linear resistivity coefficient)
follows a scaling relation with the corresponding magnitude of $T_{\rm c}$, and then
the strength of the low-temperature T-linear resistivity decreases as $T_{\rm c}$
decreases. This surprising correlation between the strength of the low-temperature
T-linear resistivity and the corresponding magnitude of $T_{\rm c}$ in the overdoped
electron-doped cuprate superconductors
\cite{Greene20,Fournier98,Jin11,Sarkar17,Sarkar19,Legros19,Yuan22}
therefore strongly suggests an intimate link between the nature of the strange
metallic normal-state and superconductivity. In other words, the origin of the
strange metallic normal-state is intertwined with the origin of superconductivity.
In this case, the strange metallic normal-state of the overdoped electron-doped
cuprate superconductors potentially serves as a starting point for a deep
understanding of the SC mechanism of cuprate superconductivity.

Although the intrinsic correlation between the strength of the low-temperature
T-linear resistivity and the corresponding magnitude of $T_{\rm c}$ in the overdoped
electron-doped cuprate superconductors has been established experimentally
\cite{Jin11,Sarkar17,Sarkar19,Legros19,Yuan22}, a complete understanding of this
correlation is still unclear. In particular, a key question is whether a common
bosonic excitation, which dominates both the electron scattering responsible for the
low-temperature T-linear resistivity and the electron pairing responsible for
superconductivity, makes a correlation between the strength of the low-temperature
T-linear resistivity and the corresponding magnitude of $T_{\rm c}$. In the recent
work \cite{Tan21}, the doping-temperature phase diagram in the electron-doped
cuprate superconductors was discussed based on the kinetic-energy-driven SC
mechanism, where
the main ingredient is identified into the constrained electron pairing mechanism
involving {\it the spin excitation, the collective mode from the internal spin
degree of freedom of the constrained electron itself}. In this case, the obtained
$T_{\rm c}$ increases with the increase of electron doping in the underdoped regime,
and reaches its maximum in the optimal electron doping, then decreases in the
overdoped regime \cite{Tan21}. On the other hand, the nature of the low-temperature
T-linear resistivity in the overdoped hole-doped cuprate superconductors was
investigated very recently \cite{Ma23}, where the electron umklapp scattering from
the spin excitation can give a consistent description of the low-temperature
T-linear resistivity. In particular, a very low temperature $T_{\rm scale}$ scales
with $\Delta^{2}_{p}$, where $\Delta_{p}$ is the minimal umklapp vector at the
antinode, and then above $T_{\rm scale}$, the resistivity is T-linear with the
strength that decreases with the increase of hole doping. In this paper, we study
the remarkable correlation between the strength of the low-temperature T-linear
resistivity and the corresponding magnitude of $T_{\rm c}$ in the overdoped
electron-doped cuprate superconductors along with these lines. We identify
explicitly that the momentum relaxation due to the electron umklapp scattering
mediated by the spin excitation reveals itself in the nature of the low-temperature
T-linear resistivity in the overdoped electron-doped cuprate superconductors as it
works in the overdoped hole-doped counterparts \cite{Ma23}. Our results in this paper
together with the previous work for superconductivity \cite{Tan21} therefore indicate
a fact: the {\it spin excitation} that mediates the attractive interaction between
the electrons responsible for superconductivity also mediates the electron umklapp
scattering responsible for the low-temperature T-linear resistivity in the overdoped
regime. This {\it same spin excitation} then generates naturally a striking
correlation between the strength of the low-temperature T-linear resistivity and the
corresponding magnitude of $T_{\rm c}$ in the overdoped electron-doped cuprate
superconductors.

This paper is organized as follows. The theoretical framework is presented in Sec.
\ref{Formalism}, where a brief review of the doping-temperature phase diagram of the
electron-doped cuprate superconductors is given for the convenience in the
discussions of the correlation between the strength of the low-temperature T-linear
resistivity and the corresponding magnitude of $T_{\rm c}$ in the overdoped
electron-doped cuprate superconductors. The transport scattering rate, however, is
obtained in terms of the spin-excitation-mediated electron umklapp scattering, and
is used to derive the resistivity within the framework of the Boltzmann transport
theory. The quantitative characteristics of the correlation between the strength of
the low-temperature T-linear resistivity and the corresponding magnitude of
$T_{\rm c}$ in the overdoped regime are presented in Section
\ref{Quantitative-characteristics}, where it is shown that as in the hole-doped
case \cite{Ma23}, the resistivity in the overdoped electron-doped cuprate
superconductors
also exhibits a crossover from the T-linear behaviour in the low-temperature region
into the quadratic in temperature (T-quadratic) behaviour in the far lower
temperature region. Finally, we give a summary in Sec. \ref{summary}. In the
Appendix \ref{Derivation-of-Tc}, we present the generalization of the main
formalisms of the kinetic-energy-driven superconductivity from the hole-doped case
to the present electron-doped case, where the detail of the calculation of the
doping dependence of $T_{\rm c}$ is given.

\section{Theoretical formalism}\label{Formalism}

\subsection{Model and electron local constraint}\label{model-constraint}

The basic element of the crystal structure of both the electron- and hole-doped
cuprate superconductors is the square-lattice copper-oxide plane in which
superconductivity occurs upon either electron or hole
doping \cite{Bednorz86,Tokura89}. As in the hole-doped case \cite{Anderson87}, the
various signature features of the electron-doped cuprate superconductors can be
also properly captured by the $t$-$J$ model on a square lattice,
\begin{eqnarray}\label{tjham}
H &=&-t\sum_{<ll'>\sigma}C^{\dagger}_{l\sigma}C_{l'\sigma}
+t'\sum_{<<ll'>>\sigma}C^{\dagger}_{l\sigma}C_{l'\sigma}\nonumber\\
&+& \mu\sum_{l\sigma}C^{\dagger}_{l\sigma}C_{l\sigma}
+J\sum_{<ll'>}{\bf S}_{l}\cdot {\bf S}_{l'},~~
\end{eqnarray}
where $C^{\dagger}_{l\sigma}$ ($C_{l\sigma}$) creates (annihilates) an electron with
spin $\sigma$ (either $\uparrow$ or $\downarrow$) on site $l$, ${\bf S}_{l}$ is a
localized spin operator with its components $S_{l}^{x}$, $S_{l}^{y}$, and
$S_{l}^{z}$, while the chemical potential $\mu$ fixes the total number of electrons.
The angle brackets $<ll'>$ and $<<ll'>>$ indicate the summations over the
nearest-neighbor (NN) and next NN pairs, respectively. The $t$-$J$ model with the NN
hopping $t$ has a particle-hole symmetry since the sign of $t$ can be absorbed by
the change of the sign of the orbital on one sublattice
\cite{Hybertsen90,Gooding94,Kim98}. However, the particle-hole asymmetry can be
properly described by the next NN hopping $t'$, which has been tested extensively
\cite{Hybertsen90,Gooding94,Kim98}. In particular, it has been shown clearly that
the asymmetry seen by the angle-resolved photoemission spectroscopy (ARPES)
observation on the hole-doped and electron-doped
cuprate superconductors is actually consistent with calculations \cite{Kim98}
performed within the $t$-$J$ model (\ref{tjham}), where all of the hopping terms
have opposite signs for the electron and hole doping, while the sign of $t'$ is of
crucial importance for the coupling of the charge motion to the spin
background \cite{Hybertsen90,Gooding94,Kim98}. Throughout this paper, the NN
magnetic exchange coupling $J$ is set as the energy unit, and $t$ and $t'$ are set
to $t/J=-2.5$ and $t'/t=0.3$, respectively, as in the previous discussions of the
low-energy electronic structure of the electron-doped cuprate superconductors
\cite{Tan21}. However, to compare with the experimental energy scales, we set
$J=1000$K.

This $t$-$J$ model (\ref{tjham}) is the strong coupling limit of the Hubbard model,
and then the crucial difficulty of its solution lies in enforcing the electron
on-site local constraint \cite{Yu92,Feng93,Zhang93,Lee06}, i.e., this $t$-$J$ model
(\ref{tjham}) is subject to a on-site local constraint of no double electron
occupancy in the hole-doped case:
$\sum_{\sigma}C^{\dagger}_{l\sigma}C_{l\sigma}\leq 1$, while it is subject a
on-site local constraint of no zero electron occupancy in the electron-doped side:
$\sum_{\sigma}C^{\dagger}_{l\sigma}C_{l\sigma}\geq 1$. In the hole-doped case, the
fermion-spin transformation \cite{Feng9404,Feng15} has been developed, where the
on-site local constraint of no double electron occupancy can be treated properly in
the actual analyses. To apply this fermion-spin transformation \cite{Feng9404,Feng15}
to the electron-doped case, the $t$-$J$ model (\ref{tjham}) in the electron
representation can be converted into the $t$-$J$ model in the hole representation
by virtue of a particle-hole transformation
$C_{l\sigma}\rightarrow f^{\dagger}_{l-\sigma}$ as \cite{Tan21},
\begin{eqnarray}\label{tjham-hole-representation}
H &=& t\sum_{<ll'>\sigma}f^{\dagger}_{l\sigma}f_{l'\sigma}
-t'\sum_{<<ll'>>\sigma}f^{\dagger}_{l\sigma}f_{l'\sigma}\nonumber\\
&-& \mu_{\rm f}\sum_{l\sigma}f^{\dagger}_{l\sigma}f_{l\sigma}
+J\sum_{<ll'>}{\bf S}_{l}\cdot {\bf S}_{l'},~~
\end{eqnarray}
where $f^{\dagger}_{l\sigma}$ ($f_{l\sigma}$) is the creation (annihilation)
operator for a hole on site $l$ with spin $\sigma$. Concomitantly, the on-site local
constraint of no zero electron occupancy in the electron representation
$\sum_{\sigma}C^{\dagger}_{l\sigma}C_{l\sigma}\geq 1$ is transformed into the
on-site local constraint of no double hole occupancy in the hole representation
$\sum_{\sigma}f^{\dagger}_{l\sigma}f_{l\sigma}\leq 1$. In this case, the $t$-$J$
model (\ref{tjham})
in both the electron- and hole-doped cases is always subject to a on-site local
constraint that double occupancy of a site by two fermions of opposite spins is not
allowed, while the difference between the electron doping and hole doping is
reflected in the sign difference of the hopping integrals as we have mentioned
above. The physics of the no double occupancy in the fermion-spin
transformation \cite{Feng9404,Feng15} is taken into account by representing the
fermion operator $f_{l\sigma}$ as a composite object created by,
\begin{eqnarray}\label{CSS}
f_{l\uparrow}=a^{\dagger}_{l\uparrow}S^{-}_{l}, ~~~~
f_{l\downarrow}=a^{\dagger}_{l\downarrow}S^{+}_{l},
\end{eqnarray}
and then the on-site local constraint of no double occupancy is satisfied in actual
analyses. The $U(1)$ gauge invariant spinful fermion operator
$a^{\dagger}_{l\sigma}=e^{i\Phi_{l\sigma}}a^{\dagger}_{l}$
($a_{l\sigma}=e^{-i\Phi_{l\sigma}}a_{l}$) creates (annihilates) a charge carrier on
site $l$, and therefore keeps track of the charge degree of freedom of the
constrained electron together with some effects of spin configuration rearrangements
due to the presence of the doped electron itself. However, the $U(1)$ gauge invariant
spin operator $S^{+}_{l}$ ($S^{-}_{l}$) keeps track of the spin degree of freedom of
the constrained electron, and therefore the collective mode from this spin degree of
freedom of the constrained electron can be interpreted as the spin excitation
responsible for the dynamical spin response of the system. In this fermion-spin
representation (\ref{CSS}), the $t$-$J$ model (\ref{tjham-hole-representation}) can
be expressed as,
\begin{eqnarray}\label{cssham}
H&=&-t\sum_{<ll'>}(a^{\dagger}_{l\uparrow}a_{l'\uparrow}S^{+}_{l'}S^{-}_{l} +a^{\dagger}_{l\downarrow}a_{l'\downarrow}S^{-}_{l'}S^{+}_{l})\nonumber\\
&+&t'\sum_{<<ll'>>}(a^{\dagger}_{l\uparrow}a_{l'\uparrow}S^{+}_{l'}S^{-}_{l} +a^{\dagger}_{l\downarrow}a_{l'\downarrow}S^{-}_{l'}S^{+}_{l})\nonumber\\
&+&\mu_{\rm a}\sum_{l\sigma}a^{\dagger}_{l\sigma}a_{l\sigma}
+J_{{\rm eff}}\sum_{<ll'>}{\bf S}_{l}\cdot {\bf S}_{l'},
\end{eqnarray}
with $J_{{\rm eff}}=(1-\delta)^{2}J$, and the doping concentration $\delta=\langle
a^{\dagger}_{l\sigma}a_{l\sigma}\rangle=\langle a^{\dagger}_{l}a_{l}\rangle$. As
in the hole-doped case \cite{Ma23}, the kinetic-energy term in the $t$-$J$ model
(\ref{cssham}) in the fermion-spin representation has been transformed into the
strong coupling between charge and spin degrees of freedom of the constrained
electron, and thus governs the essential physics in the electron-doped cuprate
superconductors.

\subsection{Kinetic-energy-driven superconductivity}\label{Effective-propagator}

In the early days of superconductivity research, the kinetic-energy-driven SC
mechanism for the hole-doped cuprate superconductors
\cite{Feng15,Feng0306,Feng12,Feng15a} was established based on the $t$-$J$ model in
the fermion-spin representation. In this kinetic-energy-driven superconductivity,
the spin-excitation-mediated attractive interaction, which pairs the charge carriers
together to form the d-wave charge-carrier pairing state, arises directly from the
coupling of the charge and spin degrees of freedom of the constrained electron in
the kinetic energy of the $t$-$J$ model. However, the electron pairs with the d-wave
symmetry are generated from this d-wave charge-carrier pairing state in terms of the
charge-spin recombination \cite{Feng15a}, and then the condensation of these electron
pairs reveals the d-wave SC ground-state. This kinetic-energy-driven SC mechanism
reveals that (i) the constrained electron has dual roles, since the glue to hold
the constrained electron pairs together is {\it the spin excitation, the collective
mode from the spin degree of freedom of the constrained electron itself}. In other
words, the constrained electrons simultaneously act to glue and to be
glued \cite{Schrieffer95,Xu23a}; (ii) the spin-excitation-mediated electron pairing
state in a way is in turn strongly influenced by the single-particle coherence,
leading to a dome-like shape doping dependence of $T_{\rm c}$. Starting from the
$t$-$J$ model (\ref{cssham}) in the fermion-spin representation, the formalism of
the kinetic-energy-driven superconductivity developed for the hole-doped case has
been generalized to the electron-doped side \cite{Tan21,Tan20}, where the
complicated line-shape in the energy distribution curve, the kink in the electron
dispersion, and the autocorrelation of the angle-resolved photoemission spectroscopy
have been discussed, and the obtained results are well consistent with the
corresponding experimental results \cite{Horio16,Song17,Lin20,Horio20b}. Our
following discussions of the correlation between the strength of the low-temperature
T-linear resistivity and the corresponding magnitude of $T_{\rm c}$ in the overdoped
electron-doped cuprate superconductors builds on the kinetic-energy-driven SC
mechanism. For the following discussions of the main physics in the core, the
details of the derivation of the main formalisms of the kinetic-energy-driven
superconductivity in the electron-doped case are presented in
Appendix \ref{Derivation-of-Tc}.
\begin{figure}[h!]
\centering
\includegraphics[scale=0.65]{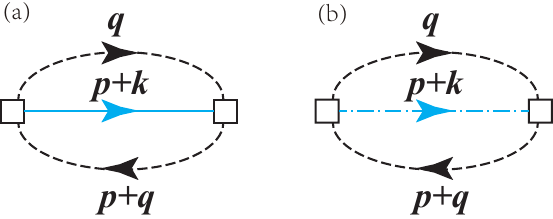}
\caption{The skeletal diagrams for the hole (a) normal and (b) anomalous
self-energies for scattering holes from the spin excitation. The blue-solid-line
and blue-dash-dot-line represent the hole diagonal and off-diagonal propagators
$G_{\rm f}$ and $\Im^{\dagger}_{\rm f}$, respectively, and the black-dash-line
depicts the spin propagator $D^{(0)}$, while $\square$ describes the vertex
function $\Lambda$. \label{hole-self-energy-diagram}}
\end{figure}
In the previous discussions \cite{Tan21,Tan20},
the full hole diagonal and off-diagonal propagators $G_{\rm f}({\bf k},\omega)$ and
$\Im^{\dagger}_{\rm f}({\bf k},\omega)$ of the $t$-$J$ model (\ref{cssham}) in the
fermion-spin representation that satisfy the self-consistent Dyson's equations have
been obtained in terms of the Eliashberg's approach \cite{Eliashberg60} as [see
Appendix \ref{Derivation-of-Tc}],
\begin{subequations}\label{HDODGF-2}
\begin{eqnarray}
G_{\rm f}({\bf k},\omega)&=&{1\over\omega-\varepsilon^{({\rm f})}_{\bf k}
-\Sigma^{(\rm f)}_{\rm tot}({\bf k},\omega)},~~~~~\label{HDGF-2}
\end{eqnarray}
\begin{eqnarray}
\Im^{\dagger}_{\rm f}({\bf k},\omega)&=&{L^{(\rm f)}({\bf k},\omega)\over\omega
-\varepsilon^{({\rm f})}_{\bf k}-\Sigma^{(\rm f)}_{\rm tot}({\bf k},\omega)},
~~~~~~~\label{HODGF-2}
\end{eqnarray}
\end{subequations}
where the non-interaction band energy
$\varepsilon^{({\rm f})}_{\bf k}=4t\gamma_{\bf k}-4t'\gamma_{\bf k}'-\mu$, with
$\gamma_{\bf k}=({\rm cos}k_{x}+{\rm cos} k_{y})/2$ and
$\gamma_{\bf k}'={\rm cos}k_{x}{\rm cos}k_{y}$,
while the total hole self-energy
$\Sigma^{(\rm f)}_{\rm tot}({\bf k},\omega)$ and the function
$L^{(\rm f)}({\bf k},\omega)$ are respectively given by,
\begin{subequations}\label{HTOTSE}
\begin{eqnarray}
\Sigma^{(\rm f)}_{\rm tot}({\bf k},\omega)&=&
\Sigma^{(\rm f)}_{\rm ph}({\bf k},\omega)
+{|\Sigma^{(\rm f)}_{\rm pp}({\bf k},\omega)|^{2}\over\omega
+\varepsilon^{(\rm f)}_{\bf k}+\Sigma^{(\rm f)}_{\rm ph}({\bf k},-\omega)},~~~~~
\end{eqnarray}
\begin{eqnarray}
L^{(\rm f)}({\bf k},\omega)&=&-{\Sigma^{(\rm f)}_{\rm pp}({\bf k},\omega)\over
\omega+\varepsilon^{(\rm f)}_{\bf k}+\Sigma^{(\rm f)}_{\rm ph}({\bf k},-\omega)}.
\end{eqnarray}
\end{subequations}

The hole normal self-energy $\Sigma^{(\rm f)}_{\rm ph}({\bf k},\omega)$ in the
particle-hole channel sketched in Fig. \ref{hole-self-energy-diagram}a and the
hole anomalous self-energy $\Sigma^{(\rm f)}_{\rm pp}({\bf k},\omega)$ in the
particle-particle channel sketched in Fig. \ref{hole-self-energy-diagram}b have
been derived in terms of the full hole diagonal and off-diagonal propagators as
[see Appendix \ref{Derivation-of-Tc}],
\begin{subequations}\label{HNASE-1}
\begin{eqnarray}
\Sigma^{(\rm f)}_{\rm ph}({\bf k},i\omega_{n})&=&{1\over N}\sum_{\bf p}
{1\over \beta}\sum_{ip_{m}}G_{\rm f}({\bf p}+{\bf k},ip_{m}+i\omega_{n})\nonumber\\
&\times& P^{(0)}({\bf k},{\bf p},ip_{m}),~~~~~~~~~~\label{HNSE-1}\\
\Sigma^{(\rm f)}_{\rm pp}({\bf k},i\omega_{n})&=&{1\over N}
\sum_{\bf p}{1\over \beta}\sum_{ip_{m}}
\Im^{\dagger}_{\rm f}({\bf p}+{\bf k},ip_{m}+i\omega_{n})\nonumber\\
&\times& P^{(0)}({\bf k},{\bf p},ip_{m}),\label{HASE-1}
\end{eqnarray}
\end{subequations}
respectively, with the number of lattice sites $N$, the fermionic and bosonic
Matsubara frequencies $\omega_{n}$ and $p_{m}$, respectively, and the effective
spin propagator,
\begin{eqnarray}\label{ESP-1}
P^{(0)}({\bf k},{\bf p},\omega)={1\over N}\sum_{\bf q}
\Lambda^{2}_{{\bf p}+{\bf q}+{\bf k}}\Pi({\bf p},{\bf q},\omega),
\end{eqnarray}
where $\Lambda_{{\bf k}}=4t\gamma_{\bf k}-4t'\gamma_{\bf k}'$ is the vertex
function. This effective spin propagator $P^{(0)}({\bf k},{\bf p},\omega)$ depict
the nature of the spin excitation, and is directly associated with the spin bubble
$\Pi({\bf p},{\bf q},\omega)$, while this spin bubble $\Pi({\bf p},{\bf q},\omega)$
is a convolution of two spin propagators, and can be expressed as,
\begin{equation}\label{spin-bubble-6}
\Pi({\bf p},{\bf q},ip_{m})={1\over\beta}\sum_{iq_{m}}D^{(0)}({\bf q},iq_{m})
D^{(0)}({\bf q}+{\bf p},iq_{m}+ip_{m}),~~~~~
\end{equation}
with the bosonic Matsubara frequency $q_{m}$, and the spin propagator
\cite{Cheng08},
\begin{eqnarray}\label{MF-spin-propagator}
D^{(0)}({\bf k},\omega)={B_{\bf k}\over \omega^{2}-\omega^{2}_{\bf k}}
={B_{\bf k}\over 2\omega_{\bf k}}\left ( {1\over\omega-\omega_{\bf k}}-
{1\over\omega+\omega_{\bf k}}\right ),~~~~~
\end{eqnarray}
where the weight function of the spin excitation spectrum $B_{{\bf k}}$ and
the spin excitation energy dispersion $\omega_{{\bf k}}$ have been given explicitly
in Appendix \ref{Derivation-of-Tc}.

With the help of the above spin propagator in Eq. (\ref{MF-spin-propagator}), the
spin bubble $\Pi({\bf p},{\bf q},\omega)$ in Eq. (\ref{spin-bubble-6}) can be
evaluated as,
\begin{eqnarray}\label{spin-bubble}
\Pi({\bf p},{\bf q},ip_{m})=-{\bar{W}^{(1)}_{{\bf p}{\bf q}}\over (ip_{m})^{2}
-[\omega^{(1)}_{{\bf p}{\bf q}}]^{2}}+{\bar{W}^{(2)}_{{\bf p}{\bf q}}\over
(ip_{m})^{2}-[\omega^{(2)}_{{\bf p}{\bf q}}]^{2}},~~~~~
\end{eqnarray}
with $\omega^{(1)}_{{\bf p}{\bf q}}=\omega_{{\bf q}+{\bf p}}+\omega_{\bf q}$,
$\omega^{(2)}_{{\bf p}{\bf q}}=\omega_{{\bf q}+{\bf p}}-\omega_{\bf q}$, and the
functions,
\begin{subequations}
\begin{eqnarray}
\bar{W}^{(1)}_{{\bf p}{\bf q}}&=&{B_{\bf q}B_{{\bf q}+{\bf p}}\over 2\omega_{\bf q}
\omega_{{\bf q}+{\bf p}}}\omega^{(1)}_{{\bf p}{\bf q}}
[n_{\rm B}(\omega_{{\bf q}+{\bf p}})+n_{\rm B}(\omega_{\bf q})+1], ~~~~~\\
\bar{W}^{(2)}_{{\bf p}{\bf q}}&=&{B_{\bf q}B_{{\bf q}+{\bf p}}\over 2\omega_{\bf q}
\omega_{{\bf q}+{\bf p}}}\omega^{(2)}_{{\bf p}{\bf q}}
[n_{\rm B}(\omega_{{\bf q}+{\bf p}})-n_{\rm B}(\omega_{\bf q})],
\end{eqnarray}
\end{subequations}
where $n_{\rm B}(\omega)$ is the boson distribution function, and then the effective
spin propagator $P^{(0)}({\bf k},{\bf p},\omega)$ in Eq. (\ref{ESP-1}) is obtained
directly from the above spin bubble in Eq. (\ref{spin-bubble}). Substituting the
effective spin propagator in Eq. (\ref{ESP-1}) into Eq. (\ref{HNASE-1}), the hole
normal and anomalous self-energies $\Sigma^{(\rm f)}_{\rm ph}({\bf k},\omega)$ and
$\Sigma^{(\rm f)}_{\rm pp}({\bf k},\omega)$ can be derived straightforwardly, and
have been given explicitly in Appendix \ref{Derivation-of-Tc}.

However, for the discussions of a link between the nature of the strange metallic
normal-state and superconductivity in the overdoped electron-doped cuprate
superconductors, we need to derive the full electron diagonal and off-diagonal
propagators $G({\bf k},\omega)$ and $\Im^{\dagger}({\bf k},\omega)$. These full
electron diagonal and off-diagonal propagators $G({\bf k},\omega)$ and
$\Im^{\dagger}({\bf k},\omega)$ are
respectively associated with the hole diagonal and off-diagonal propagators
$G_{\rm f}({\bf k},\omega)$ and $\Im^{\dagger}_{\rm f}({\bf k},\omega)$ in
Eq. (\ref{HDODGF-2}) via the particle-hole transformation
$C_{l\sigma}\rightarrow f^{\dagger}_{l-\sigma}$
as $G(l-l',t-t')=\langle\langle C_{l\sigma}(t);C^{\dagger}_{l'\sigma}(t')\rangle
\rangle=\langle\langle f^{\dagger}_{l\sigma}(t);f_{l'\sigma}(t')\rangle\rangle
=-G_{\rm f}(l'-l, t'-t)$ and
$\Im(l-l',t-t')=\langle\langle C_{l\downarrow}(t);C_{l'\uparrow}(t')\rangle
\rangle=\langle\langle f^{\dagger}_{l\uparrow}(t);f^{\dagger}_{l'\downarrow}(t')
\rangle\rangle=\Im^{\dagger}_{\rm f} (l-l',t-t')$, and have been obtained
explicitly as \cite{Tan21,Tan20},
\begin{subequations}\label{EDODGF}
\begin{eqnarray}
G({\bf k},\omega)&=&{1\over\omega-\varepsilon_{\bf k}
-\Sigma_{\rm tot}({\bf k},\omega)},~~~~~~\label{EDGF}\\
\Im^{\dagger}({\bf k},\omega)&=&{L({\bf k},\omega)\over\omega
-\varepsilon_{\bf k}-\Sigma_{\rm tot}({\bf k},\omega)},~~~~~~\label{EODGF}
\end{eqnarray}
\end{subequations}
where the electron non-interaction band energy $\varepsilon_{\bf k}
=-\varepsilon^{({\rm f})}_{\bf k}=-4t\gamma_{\bf k} +4t'\gamma_{\bf k}'+\mu$,
\begin{figure}[h!]
\centering
\includegraphics[scale=0.65]{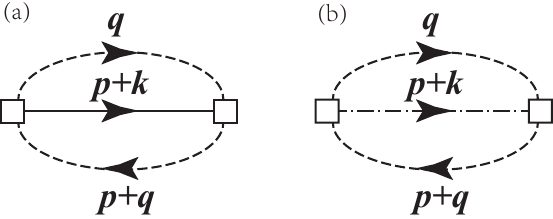}
\caption{The skeletal diagrams for the electron (a) normal and (b) anomalous
self-energies for scattering electrons from the spin excitation. The
black-solid-line and black-dash-dot-line represent the electron diagonal and
off-diagonal propagators $G$ and $\Im^{\dagger}$, respectively, and the
black-dash-line depicts the spin propagator $D^{(0)}$, while $\square$ describes
the vertex function $\Lambda$. \label{electron-self-energy-diagram}}
\end{figure}
while
the electron total self-energy $\Sigma_{\rm tot}({\bf k},\omega)$ and the function
$L({\bf k},\omega)$ can be expressed explicitly as,
\begin{subequations}\label{ETOTSE}
\begin{eqnarray}
\Sigma_{\rm tot}({\bf k},\omega)&=&\Sigma_{\rm ph}({\bf k},\omega)
+{|\Sigma_{\rm pp}({\bf k},\omega)|^{2}\over\omega+\varepsilon_{\bf k}
+\Sigma_{\rm ph}({\bf k},-\omega)},~~~~~\\
L({\bf k},\omega)&=&-{\Sigma_{\rm pp}({\bf k},\omega)\over\omega
+\varepsilon_{\bf k}+\Sigma_{\rm ph}({\bf k},-\omega)},
\end{eqnarray}
\end{subequations}
respectively, with the electron normal self-energy
$\Sigma_{\rm ph}({\bf k},\omega)$ sketched in
Fig. \ref{electron-self-energy-diagram}a and electron anomalous self-energy
$\Sigma_{\rm pp}({\bf k},\omega)$ sketched in
Fig. \ref{electron-self-energy-diagram}b that have been obtained straightforwardly
as \cite{Tan21,Tan20},
\begin{subequations}\label{ENASE}
\begin{eqnarray}
\Sigma_{\rm ph}({\bf k},\omega)&=&-\Sigma^{({\rm f})}_{\rm ph}({\bf k},-\omega),
\label{ENSE}
\end{eqnarray}
\begin{eqnarray}
\Sigma_{\rm pp}({\bf k},\omega)&=&\Sigma^{({\rm f})}_{\rm pp}({\bf k},\omega),
\label{EASE}
\end{eqnarray}
\end{subequations}
respectively. Moreover, the sharp peaks appear at low-temperature in
$\Sigma_{\rm ph}({\bf k},\omega)$, $\Sigma_{\rm pp}({\bf k},\omega)$, and
$P^{(0)}({\bf k},{\bf p},\omega)$ are actually a $\delta$-function that are
broadened by a small damping employed in the numerical calculation for a finite
lattice \cite{Brinckmann01,Restrepo23}. As the approach described in
Ref. \cite{Tan21}, the calculation in this paper for
$\Sigma_{\rm ph}({\bf k},\omega)$, $\Sigma_{\rm pp}({\bf k},\omega)$, and
$P^{(0)}({\bf k},{\bf p},\omega)$ is performed numerically on a $160\times 160$
lattice in momentum space, where the infinitesimal $i0_{+}\rightarrow i\Gamma$ is
replaced by a small damping $\Gamma=0.1J$.

\subsection{Doping dependence of $T_{\rm c}$}\label{phase-diagram}

Based on the kinetic-energy-driven SC mechanism discussed in the above subsection
\ref{Effective-propagator}, the evolution of $T_{\rm c}$ with the electron doping
in the electron-doped cuprate superconductors \cite{Tan21} has been investigated
recently by making use of the self-consistent calculation in the condition of the
SC gap $\bar{\Delta}({\bf k})=\Sigma_{\rm pp}({\bf k},0)=0$ [see Appendix
\ref{Derivation-of-Tc}].
\begin{figure}[h!]
\centering
\includegraphics[scale=0.80]{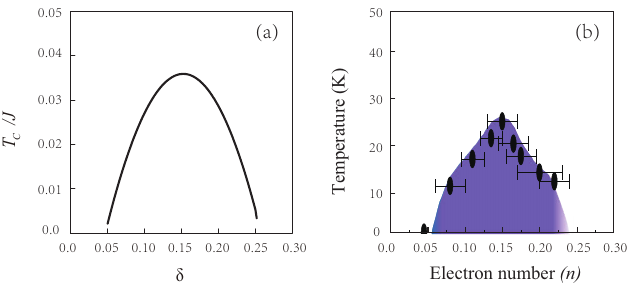}
\caption{(Color online) (a) Doping dependence of $T_{\rm c}$. (b) The corresponding
experimental result observed on Pr$_{1-x}$LaCe$_{x}$CuO$_{4-\delta}$ under the
proper annealing condition taken from Ref. \onlinecite{Song17}. \label{Tc-doping}}
\end{figure}
For the convenience in the discussions of the correlation
between the strength of the low-temperature T-linear resistivity and the
corresponding magnitude of $T_{\rm c}$ in the overdoped electron-doped cuprate
superconductors, the result of $T_{\rm c}$ as a function of the electron doping is
replotted in Fig. \ref{Tc-doping}a. For a better comparison, the corresponding
experimental result \cite{Song17} observed on the electron-doped cuprate
superconductor Pr$_{1-x}$LaCe$_{x}$CuO$_{4-\delta}$ under the proper annealing
condition is also shown in Fig. \ref{Tc-doping}b. Apparently, the experimental
result \cite{Song17} of the doping dependent $T_{\rm c}$ in
Pr$_{1-x}$LaCe$_{x}$CuO$_{4-\delta}$ under the proper annealing condition is
qualitatively reproduced, where the gradual increase in $T_{\rm c}$ with the
increase of the electron doping occurs in the underdoped regime, and $T_{\rm c}$
reaches its maximum at around the optimal electron doping $\delta\sim 0.15$,
subsequently, $T_{\rm c}$ decreases monotonically with the increase of the electron
doping in the overdoped regime.
In comparison with the corresponding result for the hole-doped case
\cite{Drozdov18,Taillefer10,Feng15a}, it thus shows that although the magnitude of
the optimized $T_{\rm c}$ in the electron-doped side is much lower than that in the
hole-doped case, the doping range of the SC dome in the electron-doped side is in a
striking analogy with that in the hole-doped case
\cite{Drozdov18,Taillefer10,Feng15a}, and in this sense, the absence of the
disparity between the phase diagrams of the electron- and hole-doped cuprate
superconductors \cite{Horio16,Song17,Lin20,Horio20b,Drozdov18,Taillefer10} is
therefore confirmed.

\subsection{Electron Fermi surface}\label{Octet-model}

The transition from the SC-state to the normal-state manifest in the anomalous
self-energy (then the full electron off-diagonal propagator) becomes zero above
$T_{\rm c}$, and then the full electron propagator in Eq. (\ref{EDODGF}) in
the SC-state is reduced in the normal-state as,
\begin{eqnarray}
G({\bf k},\omega)={1\over\omega-\varepsilon_{\bf k}-\Sigma_{\rm ph}({\bf k},\omega)},
\label{NSEGF}
\end{eqnarray}
while the electron spectrum function $A({\bf k},\omega)=-{\rm Im}G({\bf k},\omega)/\pi$
is obtained directly as,
\begin{eqnarray}\label{ESF-ED}
A({\bf k},\omega)=-{1\over\pi}{{\rm Im}\Sigma_{\rm ph}({\bf k},\omega)\over
[\omega-\varepsilon_{\bf k}-{\rm Re}\Sigma_{\rm ph}({\bf k},\omega)]^{2}
+[{\rm Im}\Sigma_{\rm ph}({\bf k},\omega)]^{2}},\nonumber\\
~~~~~
\end{eqnarray}
where ${\rm Re}\Sigma_{\rm ph}({\bf k},\omega)$ and
${\rm Im}\Sigma_{\rm ph}({\bf k},\omega)$ are the real and imaginary parts of the
electron normal self-energy $\Sigma_{\rm ph}({\bf k},\omega)$, respectively.

\begin{figure}[h!]
\centering
\includegraphics[scale=0.90]{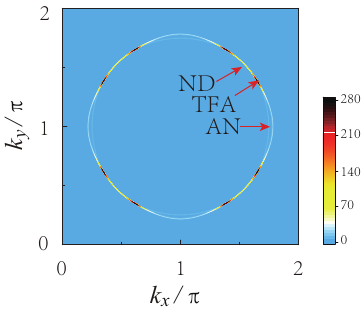}
\caption{(Color online) The map of the electron Fermi surface at $\delta=0.19$ with
$T=0.002J$, where the Brillouin zone center has been shifted by [$\pi,\pi$], and AN,
TFA, and ND denote the antinode, tip of the Fermi arc, and node, respectively.
\label{EFS-map}}
\end{figure}

The electron Fermi surface (EFS) separates the occupied and unoccupied states, and
therefore the geometrical structure of EFS is closely linked with the low-energy
electronic structure \cite{Armitage10,Campuzano04,Damascelli03,Fink07} as well as
the electrical transport \cite{Hussey08,Timusk99,Kastner98}. In the recent work
\cite{Mou17}, the geometrical structure of EFS in the electron-doped cuprate
superconductors has been studied, where the underlying EFS is obtained via the map
of the electron spectral function ({\ref{ESF-ED}}) at zero energy $\omega=0$, i.e.,
the closed EFS contour is determined by the poles of the full electron propagator
({\ref{NSEGF}}) at zero energy:
$\varepsilon_{\bf k}+{\rm Re}\Sigma_{\rm ph}({\bf k},0)=\bar{\varepsilon}_{\bf k}=0$,
with the renormalized electron energy dispersion
$\bar{\varepsilon}_{\bf k}=Z_{\rm F}\varepsilon_{\bf k}$ and the single-particle
coherent weight
$Z^{-1}_{\rm F}=1-{\rm Re}\Sigma_{\rm pho}({\bf k},0)\mid_{{\bf k}=[\pi,0]}$, while
$\Sigma_{\rm pho}({\bf k},\omega)$ that is the antisymmetric part of the electron
normal self-energy $\Sigma_{\rm ph}({\bf k},\omega)$. However, the strong
redistribution of the spectral weight at the closed EFS contour is dominated by the
the imaginary part of the electron normal self-energy
${\rm Im}\Sigma_{\rm ph}({\bf k},\omega)$ [then the single-particle scattering rate
$\Gamma_{\bf k}(\omega)=|{\rm Im}\Sigma_{\rm ph}({\bf k},\omega)|$]. To see this
point more clearly, we plot the EFS map at the electron doping $\delta=0.19$ with
temperature $T=0.002J$ in Fig. \ref{EFS-map}, where the Brillouin zone (BZ) center
has been shifted by [$\pi,\pi$], and AN, TFA, and ND indicate the antinode, tip of
the Fermi arc, and node, respectively.
Our result in Fig. \ref{EFS-map} therefore shows that EFS has been separated into
three typical regions due to the strong redistribution of the spectral weight
\cite{Mou17}: (i) the antinodal region, where the spectral weight is reduced
strongly, leading to EFS at around the antinodal region to become invisible; (ii)
the nodal region, where the spectral weight is reduced moderately, leading to EFS
to be clearly visible as the reminiscence of the EFS contour in the case of the
absence of the electron interaction to form the Fermi arcs; (iii) the region at
around the tips of the Fermi arcs, where the spectral weight exhibits the largest
value. This EFS reconstruction is also qualitatively consistent with the recent
experimental observations on the electron-doped cuprate superconductors
\cite{Horio16,Song17,Lin20,Horio20b}, where upon the optimal annealing, the
weight of the ARPES spectrum around the antinodal region is reduced, and then
EFS is truncated to form the Fermi arcs located around the nodal region.

\begin{figure}[h!]
\centering
\includegraphics[scale=0.90]{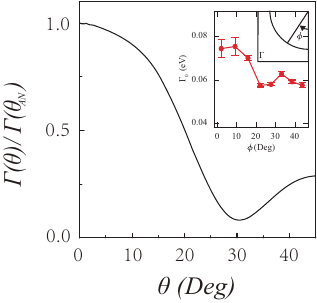}
\caption{The single-particle scattering rate
$\Gamma(\theta)/\Gamma(\theta_{\rm AN})$ as a function of Fermi angle $\theta$ at
$\delta=0.19$ with $T=0.002J$ for $\omega=0$, where $\Gamma(\theta_{\rm AN})$ is
the single-particle scattering rate at the antinode. Inset: the corresponding
experimental result of the overdoped Pr$_{1.3-x}$La$_{0.7}$Ce$_{x}$CuO$_{4}$ taken
from Ref. \onlinecite{Horio16}.\label{single-particle-scattering}}
\end{figure}

In the previous studies \cite{Mou17}, it has been shown that the spectral
redistribution to form the Fermi arcs is directly associated with the angular
dependence of the single-particle scattering rate $\Gamma(\theta)$, where
$\Gamma(\theta)=\Gamma_{{\rm k}_{\rm F}(\theta)}(0)
=|{\rm Im}\Sigma_{\rm ph}({\rm k}_{\rm F}(\theta),0)|$ with the Fermi angle
$\theta$. To see this $\Gamma(\theta)$ in momentum space more clearly, we plot the
$\Gamma(\theta)/\Gamma(\theta_{\rm AN})$ along EFS from the antinode to the node at
$\delta=0.19$ with $T=0.002J$ in Fig. \ref{single-particle-scattering}, where
$\Gamma(\theta_{\rm AN})$ is the single-particle scattering rate at the antinode.
For comparison, the corresponding experimental result of the single-particle
scattering rate along EFS observed from the overdoped electron-doped cuprate
superconductor \cite{Horio16} Pr$_{1.3-x}$La$_{0.7}$Ce$_{x}$CuO$_{4}$ is also
presented in Fig. \ref{single-particle-scattering} (inset).
It thus shows that the
actual minimum of $\Gamma(\theta)$ does not site at the node, but appears at around
the tip of the Fermi arc. On the other hand, the magnitude of $\Gamma(\theta)$ still
exhibits the largest value at the antinode, and then it decreases when the momentum
moves away from the antinode. Moreover, the magnitude of $\Gamma(\theta)$ at the
antinode is much larger than that at the node. This special angular dependence of
$\Gamma(\theta)$ therefore generates a spectral redistribution to form the Fermi arcs
with the largest value of the spectral weight located at around the tips of the Fermi
arcs.

\subsection{Momentum relaxation in the normal-state }\label{Boltzmann-theory}

We now turn to discuss the normal-state electrical transport in the overdoped
electron-doped cuprate superconductors. As in the hole-doped case
\cite{Campuzano04,Damascelli03,Fink07}, the conventional quasiparticle picture
breaks down in the normal-state of the overdoped electron-doped cuprate
superconductors \cite{Armitage10}. In this case, two different methods have been
employed to investigate the electrical transport in the case of the
lack of the well-defined quasiparticle. The first one is the memory-matrix
transport approach \cite{Mahajan13,Hartnoll14,Patel14,Lucas15,Vieira20,Mandal21},
which has a distinct advantage of not relying on the existence of the well-defined
quasiparticle. In particular, the electrical transports in the strange-metal phases
of different strongly correlated systems have been investigated based on the
memory-matrix transport approach
\cite{Mahajan13,Hartnoll14,Patel14,Lucas15,Vieira20,Mandal21}, and the obtained
results are consistent with the corresponding experimental data. The another method
employed is the Boltzmann transport theory \cite{Abrikosov88,Mahan81}, where it is
crucial to assume either the validity of the conventional quasiparticle picture or
the treatment of the electron interaction mediated by different bosonic modes within
the Eliashberg approach \cite{Prange64,Lee21}. This follows a basic fact that (i) in
the early pioneering work \cite{Prange64}, Prange and Kadanoff have shown that in an
electron-phonon system, a set of transport equations can be derived in the Migdal's
approximation. In particular, this coupled set of transport equations for the
electron and phonon distribution functions is correct even in the case in which the
electron excitation spectrum has considerable width and structure so that one might
not expect a priori that there would be the well-defined quasiparticle
\cite{Prange64}. Nevertheless, one of the forms of the electrical transport equation,
\begin{eqnarray}\label{Boltzmann-equation-2}
e{\bf E}\cdot\nabla_{\bf k}f({\bf k})=I_{\rm e-e},
\end{eqnarray}
is identical to the electrical transport equation proposed by Landau for the case
of the existence of the well-defined quasiparticle \cite{Abrikosov88,Mahan81}, where
$e$ is the charge of an electron, and $f({\bf k},t)$ is the distribution function
in a homogeneous system.
For the convenience in the following discussions, the
external magnetic field ${\bf H}$ has been ignored, and only an external electric
field ${\bf E}$ is applied to the system, while $I_{\rm e-e}$ is the
electron-electron collision term, and is directly related to the momentum relaxation
mechanism; (ii) More importantly, it has been confirmed recently that this transport
equation (\ref{Boltzmann-equation-2}) developed by Prange and Kadanoff
\cite{Prange64}
is not specific to a phonon-mediated interaction, and also is valid for the system
with the interaction mediated by other bosonic excitations \cite{Lee21}.

In the subsequent analysis, we investigate the low-temperature resistivity in the
overdoped electron-doped cuprate superconductors based on the Boltzmann transport
equation (\ref{Boltzmann-equation-2}). To derive this Boltzmann transport equation
(\ref{Boltzmann-equation-2}), the linear perturbation from the equilibrium in terms
of the distribution function,
\begin{eqnarray}\label{distribution-function}
f({\bf k})&=&n_{\rm F}({\bar{\varepsilon}_{\bf k}})
-{d n_{\rm F}({\bar{\varepsilon}_{\bf k}})
\over d{\bar{\varepsilon}_{\bf k}}}\tilde{\Phi}({\bf k}),
\end{eqnarray}
can be introduced as it has been done in the previous works
\cite{Ma23,Prange64,Lee21}, where $n_{\rm F}(\omega)$ is the fermion distribution
function, $\tilde{\Phi}({\bf k})$ is a local shift of the
chemical potential at a given patch of EFS \cite{Ma23,Prange64,Lee21}, and obeys the
antisymmetric relation $\tilde{\Phi}(-{\bf k})=-\tilde{\Phi}({\bf k})$. Substituting
the above result in Eq. (\ref{distribution-function}) into
Eq. (\ref{Boltzmann-equation-2}), we can linearize the Boltzmann equation
(\ref{Boltzmann-equation-2}) as,
\begin{eqnarray}\label{Boltzmann-equation-3}
e{\bf v}_{\bf k}\cdot{\bf E}{dn_{\rm F}({\bar{\varepsilon}_{\bf k}})\over
d{\bar{\varepsilon}_{\bf k}}}=I_{\rm e-e},
\end{eqnarray}
with the electron velocity
${\bf v}_{\bf k}=\nabla_{\bf k}{\bar{\varepsilon}_{\bf k}}$.

\begin{figure}[h!]
\centering
\includegraphics[scale=0.80]{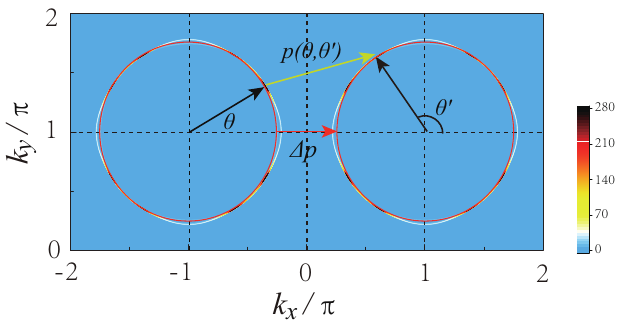}
\caption{(Color online) Schematic picture of the electron umklapp scattering process
\cite{Lee21}, where an electron on a electron Fermi surface (left) is scattered by
its partner on the umklapp electron Fermi surface (right). The intensity map of the
electron Fermi surface is the same as shown in Fig. \ref{EFS-map}, where the Fermi
wave vector of the tips of the Fermi arcs ${\rm k}^{\rm TFA}_{\rm F}$ is the radius
of the circular electron Fermi surface (red), and then an electron on this circular
electron Fermi surface (left) parametrized by the Fermi angle $\theta$ is scattered
to a point parametrized by the Fermi angle $\theta'$ on the umklapp electron Fermi
surface (right) by the spin excitation carrying momentum ${\rm p}(\theta,\theta')$.
$\Delta_{p}$ is the minimal umklapp vector at the antinode (the Fermi angle
$\theta=0$).\label{scatter-process}}
\end{figure}

We now focus on the electron-electron collision term $I_{\rm e-e}$, which is closely
associated with the momentum relaxation mechanism \cite{Abrikosov88,Mahan81}.
Although the momentum relaxation mechanism underlying the low-temperature T-linear
resistivity in the strange metallic normal-state still remains controversial to
date, it is likely of the electronic origin. In particular, it has been shown
clearly that the electron umklapp scattering plays a crucial role in the
normal-state transport of cuprate superconductors \cite{Lee21,Rice17,Hussey03}. In
this paper, we employ the electron umklapp scattering to study the low-temperature
resistivity in the overdoped electron-doped cuprate superconductors.
To see this electron umklapp scattering process more clearly, a schematic picture
of the electron umklapp scattering process \cite{Lee21} is shown in
Fig. \ref{scatter-process}, where an electron on a circular EFS (left) is scattered
by its partner on the umklapp EFS (right). The intensity map of EFS in
Fig. \ref{scatter-process} is identical to that shown in Fig. \ref{EFS-map}, where
the Fermi wave vector of the tips of the Fermi arcs ${\rm k}^{\rm TFA}_{\rm F}$ is
the radius of the circular EFS (red). This circle EFS (red) connects all tips of the
Fermi arcs, and then almost all the fraction of the spectral weight is located on
this circular EFS.

In the recent discussions \cite{Lee21}, it has been shown that the electrical
transport in the strange metallic normal-state of the overdoped hole-doped cuprate
superconductors arises from the umklapp scattering between electrons by the exchange
of a critical boson propagator, where the anisotropic transport scattering rate is
T-linear in the low-temperature region near the umklapp point, which induces a
low-temperature T-linear resistivity. Very recently, we \cite{Ma23} have also studied
the nature of the electrical transport in the strange metallic normal-state of the
overdoped hole-doped cuprate superconductors, where the momentum dependence of the
transport scattering rate originates from the umklapp scattering between electrons
by the exchange of the effective spin propagator, and scales linearly with
temperature in the low-temperature region, which then naturally generates a
low-temperature T-linear resistivity. Following these recent discussions
\cite{Ma23,Lee21}, the electron-electron collision $I_{\rm e-e}$ in
Eq. (\ref{Boltzmann-equation-3}) in the present case can be derived as,
\begin{widetext}
\begin{eqnarray}\label{electron-collision-1}
I_{\rm e-e}&=&{1\over N^{2}}\sum_{{\bf k}',{\bf p}} {2\over T}
|P({\bf k},{\bf p},{\bf k}',\bar{\varepsilon}_{\bf k}
-\bar{\varepsilon}_{{\bf k}+{\bf p}+{\bf G}})|^{2}
\{\tilde{\Phi}({\bf k})+\tilde{\Phi}({\bf k'})
-\tilde{\Phi}({\bf k}+{\bf p}+{\bf G})-\tilde{\Phi}({\bf k}'-{\bf p})\}\nonumber\\
&\times& n_{\rm F}(\bar{\varepsilon}_{\bf k})
n_{\rm F}(\bar{\varepsilon}_{{\bf k}'})
[1-n_{\rm F}(\bar{\varepsilon}_{{\bf k}+{\bf p}+{\bf G}})]
[1-n_{\rm F}(\bar{\varepsilon}_{{\bf k}'-{\bf p}})]
\delta(\bar{\varepsilon}_{\bf k}+\bar{\varepsilon}_{\bf k'}
-\bar{\varepsilon}_{{\bf k}+{\bf p}+{\bf G}}-\bar{\varepsilon}_{{\bf k}'-{\bf p}}),
\end{eqnarray}
\end{widetext}
where ${\bf G}$ labels a set of reciprocal lattice vectors, and then the above
electron umklapp scattering (\ref{electron-collision-1}) is described as a
scattering between electrons by the exchange of the effective spin propagator,
\begin{eqnarray}\label{ESP}
P({\bf k},{\bf p},{\bf k}',\omega)&=&{1\over N}\sum_{\bf q}
\Lambda_{{\bf p}+{\bf q}+{\bf k}}\Lambda_{{\bf q}+{\bf k}'}
\Pi({\bf p},{\bf q},\omega),~~~
\end{eqnarray}
rather than the scattering between electrons via the emission and absorption of the
spin excitation \cite{Ma23}.

The electron-electron collision $I_{\rm e-e}$ in Eq. (\ref{electron-collision-1})
are both functions of momentum and energy, where the magnitude of wave vector
dependence is unimportant at low temperatures, since everything happens at EFS
\cite{Abrikosov88,Mahan81,Prange64,Lee21,Rice17,Hussey03,Haldane18}. In this case,
a given patch at the circular EFS shown in Fig. \ref{scatter-process} is depicted
via the Fermi angle $\theta$ with the angle range $\theta\in [0,2\pi]$, which leads
to that the momentum integration along the perpendicular momentum is replaced by
the integration over $\bar{\varepsilon}_{\bf k}$ \cite{Prange64,Lee21}.
In the present case of the umklapp scattering between electrons by the exchange of
the effective spin propagator, an electron on the circular EFS parametrized by the
Fermi angle $\theta$ is scattered to a point parametrized by the Fermi angle
$\theta'$ on the umklapp EFS in terms of the spin excitation carrying momentum
${\rm p}(\theta,\theta')$ as shown in Fig. \ref{scatter-process}. With the help of
the above treatment, the electron-electron collision $I_{\rm e-e}$ in
Eq. (\ref{electron-collision-1}) can be derived straightforwardly, and has been
given explicitly in Ref. \onlinecite{Ma23}. In this case, the Boltzmann transport
equation (\ref{Boltzmann-equation-3}) can be obtained as,
\begin{eqnarray}\label{electron-collision}
e{\bf v}_{\rm F}(\theta)\cdot {\bf E}=-2\int {d\theta'\over {2\pi}}\zeta(\theta')
F(\theta,\theta')[\Phi(\theta)-\Phi(\theta')],~~~~~
\end{eqnarray}
with $\Phi(\theta)=\tilde{\Phi}[{\rm k}(\theta)]$, the Fermi velocity
${\bf v}_{\rm F}(\theta)$ at the Fermi angle $\theta$, the density of states factor
$\zeta(\theta')={\rm k}^{2}_{\rm F}/[4\pi^{2}{\rm v}^{3}_{\rm F}]$ at angle
$\theta'$, the Fermi wave vector ${\rm k}_{\rm F}$, and the Fermi velocity
${\rm v}_{\rm F}$. In this case, the antisymmetric relation
$\tilde{\Phi}(-{\bf k})=-\tilde{\Phi}({\bf k})$ for $\tilde{\Phi}({\bf k})$ in
Eq. (\ref{distribution-function}) is replaced by the antisymmetric relation
$\Phi(\theta)=-\Phi(\theta+\pi)$ for $\Phi(\theta)$ in the above
Eq. (\ref{electron-collision}). Moreover, the coefficient of $\Phi(\theta)$ in the
first term of the right-hand side of Eq. (\ref{electron-collision}),
\begin{eqnarray}\label{scattering-rate}
\gamma(\theta)=2\int {d \theta' \over {2\pi}} \zeta(\theta')F(\theta,\theta'),
\end{eqnarray}
is identified as the angular dependence of the transport scattering rate
\cite{Ma23,Lee21}, where as shown in Fig. \ref{scatter-process}, the kernel function
$F(\theta,\theta')$ links up the point $\theta$ on the circular EFS with the point
$\theta'$ on the umklapp EFS via the magnitude of the momentum transfer
${\rm p}(\theta,\theta')$, and can be expressed explicitly as,
\begin{eqnarray}\label{kernel-function}
F(\theta,\theta')&=&{1\over T}\int {d\omega\over 2\pi}{\omega^{2}\over
{\rm p}(\theta,\theta')}
{|\bar{P}[{\rm k}(\theta),{\rm p}(\theta,\theta'),\omega]|}^{2}\nonumber\\
&\times& n_{\rm B}(\omega)[1+n_{\rm B}(\omega)],~~~~~~
\end{eqnarray}
where the reduced effective spin propagator
$\bar{P}[{\rm k}(\theta),{\rm p}(\theta,\theta'),\omega]$ has been given
explicitly in Ref. \cite{Ma23}.

\section{Scaling of low-temperature transport scattering}
\label{Quantitative-characteristics}

The electron current density now can be derived via the local shift of the chemical
potential $\Phi(\theta)$ as \cite{Ma23},
\begin{eqnarray}\label{current-density}
{\bf J} &=& en_{0}{1\over N}\sum_{\bf k}{\bf v}_{\bf k}
{dn_{\rm F}({\bar{\varepsilon}_{\bf k}})\over d\bar{\varepsilon}_{\bf k}}
\tilde{\Phi}({\bf k})\nonumber\\
&=& -en_{0}{{\rm k}_{\rm F}\over {\rm v}_{\rm F}}\int
{d\theta\over (2\pi)^{2}}{\bf v}_{\rm F}(\theta)\Phi(\theta),~~~~~
\end{eqnarray}
where the momentum relaxation is induced by the action of the electric field on the
mobile electrons at EFS with the density $n_{0}$. In particular, the
local shift of the chemical potential $\Phi(\theta)$ can be obtained in the
relaxation-time approximation as \cite{Ma23,Lee21},
$\Phi(\theta)=-e{\rm v}_{\rm F}{\rm cos}(\theta)E_{\hat{x}}/[2\gamma(\theta)]$, with
the electric field ${\bf E}$ that has been chosen along the $\hat{x}$-axis. In this
case, the dc conductivity can be obtained straightforwardly as \cite{Ma23,Lee21},
\begin{eqnarray}\label{dc-conductivity}
\sigma_{\rm dc}(T)={1\over 2}e^{2}n_{0}{\rm k}_{\rm F}{\rm v}_{\rm F}
\int {d\theta\over (2\pi)^{2}}{\rm cos}^{2}(\theta)
{1\over\gamma(\theta)},
\end{eqnarray}
while the resistivity is related directly to the above dc conductivity,
and can be expressed explicitly as,
\begin{eqnarray}\label{dc-resistivity}
\rho(T)={1\over \sigma_{\rm dc}(T)}.
\end{eqnarray}
The above results therefore show that the electron-electron collision $I_{\rm e-e}$
(\ref{electron-collision-1}) originated from the umklapp scattering between
electrons by the exchange of the effective spin propagator leads to the appearance
of the electrical resistance.

\begin{figure}[h!]
\centering
\includegraphics[scale=0.90]{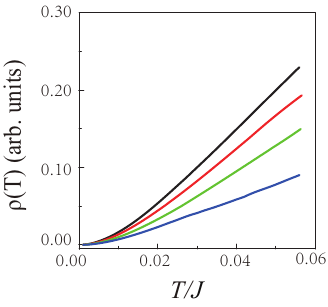}
\caption{(Color online) The resistivity as a function of temperature at
$\delta=0.15$ (black-line), $\delta=0.17$ (red-line), $\delta=0.19$ (green-line),
and $\delta=0.21$ (blue-line).\label{resistivity-temperature}}
\end{figure}

\begin{figure}[h!]
\centering
\includegraphics[scale=0.75]{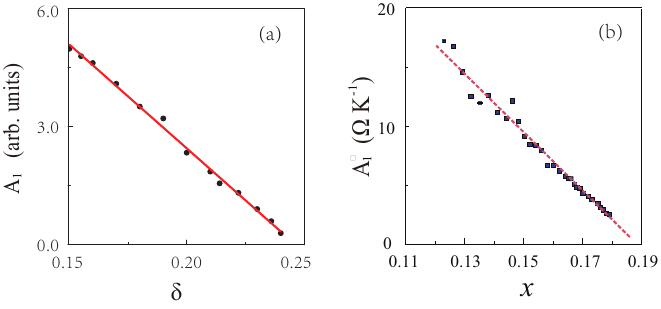}
\caption{(a) The strength of the low-temperature T-linear resistivity (black dots)
as a function of doping. (b) The corresponding experimental result of
La$_{2-x}$Ce$_{x}$CuO$_{4}$ taken from Ref. \onlinecite{Yuan22}.
\label{coefficient-doping}}
\end{figure}

Now we are ready to discuss the exotic properties of the low-temperature resistivity
in the overdoped electron-doped cuprate superconductors.
In Fig. \ref{resistivity-temperature}, we plot the resistivity $\rho(T)$ in
Eq. (\ref{dc-resistivity}) as a function of temperature at the electron doping
concentrations $\delta=0.15$ (black-line), $\delta=0.17$ (red-line), $\delta=0.19$
(green-line), and $\delta=0.21$ (blue-line), where the temperature region is clearly
divided into two characteristic regions: (i) the low-temperature region
($T>0.01J\approx 10$K), where the resistivity $\rho(T)$ is T-linear over a wide range
of the electron doping in the overdoped regime, with the strength of the T-linear
resistivity (then the T-linear resistivity coefficient) $A_{1}$ that decreases with
the increase of the electron doping all the way up to the edge of the SC dome, in
qualitative agreement with the corresponding experimental
results \cite{Greene20,Fournier98,Jin11,Sarkar17,Sarkar19,Legros19,Yuan22}.
To see this evolution of the T-linear resistivity strength $A_{1}$ with the electron
doping concentration more clearly, we plot $A_{1}$ as a function of the electron
doping in Fig. \ref{coefficient-doping}a, where the data (black dots) are the
numerical results from Eq. (\ref{dc-resistivity}), while the red-line is guide for
the eyes. For a better comparison, the corresponding experimental result
\cite{Yuan22}
observed on the electron-doped cuprate superconductor La$_{2-x}$Ce$_{x}$CuO$_{4}$ in
the overdoped regime is also shown in Fig. \ref{coefficient-doping}b. The obtained
result in Fig. \ref{coefficient-doping}a thus shows that $A_{1}$ almost linearly
{\it decreases} with the increase of the electron doping concentration for the
overdoping up to the edge of the SC dome, also in qualitative agreement with the
corresponding experimental data
\cite{Greene20,Fournier98,Jin11,Sarkar17,Sarkar19,Legros19,Yuan22}; (ii) the
far-lower-temperature region ($T< 0.01J\approx 10$K), where the resistivity
$\rho(T)$ is nonlinear in temperature. To see this nonlinear behaviour more
clearly, we have fitted the present result of the resistivity $\rho(T)$ in the
far-lower-temperature region, and found that in the far-lower-temperature region,
the resistivity decreases quadratically with the decrease of temperature.

The low-temperature T-linear resistivity occurring in the overdoped electron-doped
cuprate superconductors
\cite{Greene20,Fournier98,Jin11,Sarkar17,Sarkar19,Legros19,Yuan22} has been also
detected experimentally in the overdoped hole-doped cuprate superconductors
\cite{Daou09,Cooper09,Ayres21,Grisso21}, indicating that the
low-temperature T-linear resistivity is a generic feature in the strange-metal
phase of both the overdoped electron- and hole-doped cuprate superconductors.
These experimental observations
\cite{Greene20,Fournier98,Jin11,Sarkar17,Sarkar19,Legros19,Yuan22,Daou09,Cooper09,Ayres21,Grisso21}
show the same physical origin of the low-temperature T-linear resistivity in both
the overdoped electron- and hole-doped cuprate superconductors. In this case, we
\cite{Ma23} have also investigated the low-temperature T-linear resistivity in
the overdoped hole-doped cuprate superconductors, and the results show that as in
the present case of the low-temperature T-linear resistivity in the electron-doped
cuprate superconductors, the mechanism
of the momentum relaxation for the low-temperature T-linear resistivity in the
overdoped hole-doped cuprate superconductors originates from the same umklapp
scattering between electrons by the exchange of the effective spin propagator.

In addition to the above results of the low-temperature T-linear
resistivity, we have also performed a numerical calculation for the transport
scattering rate $\gamma(\theta)$ in Eq. (\ref{scattering-rate}), and the obtained
results show that the overall feature of the strong angular dependence of
$\gamma(\theta)$ in the electron-doped side is quite similar to the corresponding
one in the hole-doped case \cite{Ma23}, where $\gamma(\theta)$ has the largest
value at around the antinodal region, and then it decreases with the move of the
Fermi angle away from the antinode. In particular, although the magnitude of
$\gamma(\theta)$ at around the nodal region is smaller than that at around the
antinodal region, $\gamma(\theta)$ exhibits its minimum at around the tips of the
Fermi arcs, which therefore show that the normal-state transport is mainly governed
by $\gamma(\theta)$ at the antinodal region.
\begin{figure}[h!]
\centering
\includegraphics[scale=0.75]{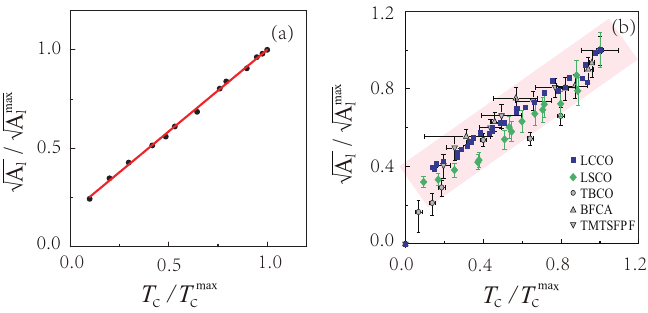}
\caption{(a) The correlation (black dots) between the square-root of the T-linear
resistivity strength $\sqrt{A_{1}}/\sqrt{A^{\rm max}_{1}}$ and the corresponding
magnitude of $T_{\rm c}/T^{\rm max}_{\rm c}$, where $A^{\rm max}_{1}$ and
$T^{\rm max}_{\rm c}$ are maximal values of the T-linear resistivity strength and
superconducting transition temperature at the optimal electron doping, respectively.
(b) The experimental results taken from Ref. \onlinecite{Yuan22}, where the squares are
the data from La$_{2-x}$Ce$_{x}$CuO$_{4}$, the diamonds are the data from
La$_{2-x}$Sr$_{x}$CuO$_{4}$, the circles are the data from
Tl$_{2}$Ba$_{2}$CuO$_{6+\delta}$, the upper triangles are the data from
Ba(Fe$_{1-x}$Co$_{x}$)$_{2}$As$_{2}$, and the lower triangles are the data from
(TMTSF)$_{2}$PF$_{6}$. \label{A1-Tc}}
\end{figure}
This also follows a fact that as in
the hole-doped case \cite{Ma23}, the characteristic feature of the tips of the
Fermi arcs shown in Fig. \ref{EFS-map} is that both the real and
imaginary parts of the electron normal self-energy have the anomalously small values
\cite{Tan21,Mou17}, indicating that the interaction (then the scattering) between
electrons at around the tips of the Fermi arcs is particularly weak. In other words,
although the electron density of states is largest at around the tips of the Fermi
arcs, the electron scattering at around the tips of the Fermi arcs is quite weak,
and then the electrons at around the tips of the Fermi arcs move more freely than
those at other parts of EFS. The current results together with our recent work for
the hole-doped case \cite{Ma23} therefore also show that the electron umklapp
scattering from the spin excitation can give a consistent description of the
low-temperature T-linear resistivity in both the overdoped hole- and
electron-doped cuprate superconductors.

We now turn to discuss the intrinsic correlation between the strength of the
low-temperature T-linear resistivity and the corresponding magnitude of $T_{\rm c}$
in the overdoped electron-doped cuprate superconductors. In Fig. \ref{A1-Tc}a, we
plot the square-root of the strength of the low-temperature T-linear resistivity
$\sqrt{A_{1}}$ as a function of $T_{\rm c}$. The theoretical results (black dots)
in Fig. \ref{A1-Tc}a are extracted from the data in Fig. \ref{Tc-doping}a and
Fig. \ref{coefficient-doping}a, while the red-line is guide for the eyes. For a
clear comparison, the corresponding experimental result \cite{Yuan22} observed on
La$_{2-x}$Ce$_{x}$CuO$_{4}$ in the overdoped regime is also shown in
Fig. \ref{A1-Tc}b (the squares). It thus shows clearly in Fig. \ref{A1-Tc}
that the experimental result \cite{Yuan22} of the scaling relation between the
strength of the low-temperature T-linear resistivity $\sqrt{A_{1}}$ and the
corresponding magnitude
of $T_{\rm c}$ in the overdoped electron-doped cuprate superconductors is
qualitatively reproduced. $\sqrt{A_{1}}/\sqrt{A^{\rm max}_{1}}$ is grown linearly
when $T_{\rm c}/T^{\rm max}_{\rm c}$ is raised, with $A^{\rm max}_{1}$ and
$T^{\rm max}_{\rm c}$ that are maximal values of the strength of the T-linear
resistivity and SC transition temperature at the optimal electron doping,
respectively.
In other words, the evolution of $T_{\rm c}$ correlates with the
emergence of the low-temperature T-linear resistivity in the overdoped regime, and
then the magnitude of $\sqrt{A_{1}}$ decreases as $T_{\rm c}$ decreases. The
present result of the scaling relation in Fig. \ref{A1-Tc} therefore also confirms
the intimate connection between the nature of the strange metallic normal-state
and superconductivity observed on the overdoped electron-doped cuprate
superconductors \cite{Greene20,Fournier98,Jin11,Sarkar17,Sarkar19,Legros19,Yuan22}.
It should be noted that this correlation between the strength of the
low-temperature T-linear resistivity and the corresponding magnitude of $T_{\rm c}$
may be a universal feature of not only the cuprate superconductors but also many
other unconventional superconductors, since as the experimental data \cite{Yuan22}
shown in Fig. \ref{A1-Tc}b, the correlation between the strength of the
low-temperature T-linear resistivity and the corresponding magnitude of $T_{\rm c}$
has been observed experimentally on the overdoped hole-doped cuprate superconductors
\cite{Taillefer10,Yuan22,Bozovic16}, such as La$_{2-x}$Sr$_{x}$CuO$_{4}$ (the
diamonds in Fig. \ref{A1-Tc}b) and Tl$_{2}$Ba$_{2}$CuO$_{6+\delta}$ (the circles in
Fig. \ref{A1-Tc}b), and many other unconventional superconductors
\cite{Taillefer10,Yuan22,Leyraud09}, such as the single-band organic superconductor
(TMTSF)$_{2}$PF$_{6}$ (the lower triangles in Fig. \ref{A1-Tc}b) as well as the
iron-based superconductor Ba(Fe$_{1-x}$Co$_{x}$)$_{2}$As$_{2}$ (the upper
triangulars in Fig. \ref{A1-Tc}b).

The essential physics of the crossover from the T-linear resistivity in the
low-temperature region into T-quadratic resistivity in the far-lower-temperature
region invoked for the overdoped hole-doped case \cite{Ma23} straightforwardly
applies here on the overdoped electron-doped side, and is attributed to the
electron umklapp scattering mediated by {\it the spin excitation}. The nature of
the transport scattering rate in Eq. (\ref{scattering-rate}) [then the nature of
the resistivity in Eq. (\ref{dc-resistivity})] is dominated by the
nature of the kernel function $F(\theta,\theta')$ in Eq. (\ref{kernel-function}).
However, this kernel function $F(\theta,\theta')$ is proportional to the effective
spin propagator $P({\bf k},{\bf p},{\bf k}',\omega)$ in Eq. (\ref{ESP}), which can
be expressed more clearly as,
\begin{eqnarray}\label{reduced-propagator}
P({\bf k},{\bf p},{\bf k}',\omega)&=&-{1\over N}\sum\limits_{\bf q}\left [
{\varpi_{1}({\bf k},{\bf p},{\bf k}',{\bf q})\over\omega^{2}
-[\omega^{(1)}_{{\bf p}{\bf q}}]^{2}}\right . \nonumber\\
&-& \left . {\varpi_{2}({\bf k},{\bf p},{\bf k}',{\bf q})\over\omega^{2}
-[\omega^{(2)}_{{\bf p}{\bf q}}]^{2}} \right ],~~~~~
\end{eqnarray}
with the functions,
\begin{subequations}
\begin{eqnarray}
\varpi_{1}({\bf k},{\bf p},{\bf k}',{\bf q})&=& \Lambda_{{\bf k}+{\bf p}+{\bf q}}
\Lambda_{{\bf q}+{\bf k}'}\bar{W}^{(1)}_{{\bf p}{\bf q}},\\
\varpi_{2}({\bf k},{\bf p},{\bf k}',{\bf q})&=& \Lambda_{{\bf k}+{\bf p}+{\bf q}}
\Lambda_{{\bf q}+{\bf k}'}\bar{W}^{(2)}_{{\bf p}{\bf q}}.
\end{eqnarray}
\end{subequations}
\begin{figure}[h!]
\centering
\includegraphics[scale=0.95]{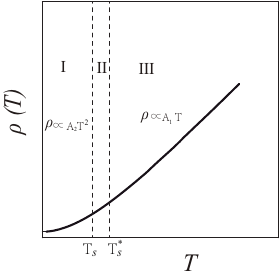}
\caption{Schematic regions of the low-temperature resistivity in the overdoped
electron-doped cuprate superconductors, where
$T^{*}_{\rm s}=T_{\rm scale}+\omega_{{\bf k}_{\rm A}}$, $\omega_{{\bf k}_{\rm A}}$
is the spin excitation energy at the $[\pi,\pi]$ point of the Brillouin zone, and
$T_{\rm s}=T_{\rm scale}$. (I) The far-lower-temperature region ($T<T_{\rm s}$),
where the resistivity decreases quadratically as the temperature decreases; (II)
The extremely-narrow-crossover region ($T_{\rm s}<T<T^{*}_{\rm s}$), where the
resistivity is neither T-linear nor T-quadratic, but is a nonlinear in temperature;
(III) The low-temperature region ($T>T^{*}_{\rm s}$), where the resistivity is
T-linear.\label{three-region}}
\end{figure}
Following the recent discussions for the hole-doped case \cite{Ma23}, we can also
make a taylor expansion for the effective spin excitation energy dispersions
$\omega^{(1)}_{{\bf p}{\bf q}}$ and $\omega^{(2)}_{{\bf p}{\bf q}}$ in the
electron-doped side, and then the effective spin excitation energy dispersions
$\omega^{(1)}_{{\bf p}{\bf q}}$ and $\omega^{(2)}_{{\bf p}{\bf q}}$ can be obtained
approximately as,
\begin{subequations}\label{effective-spin-excitation}
\begin{eqnarray}
\omega^{(1)}_{{\bf p}{\bf q}}&=&\omega_{{\bf q}+{\bf p}}
+\omega_{\bf q}\approx a({\bf q})p^{2}+
2\omega_{\bf q},
\end{eqnarray}
\begin{eqnarray}
\omega^{(2)}_{{\bf p}{\bf q}}&=&\omega_{{\bf q}+{\bf p}}
-\omega_{\bf q}\approx a({\bf q})p^{2},
\end{eqnarray}
\end{subequations}
with $a({\bf q})=(d^{2}\omega_{\bf q}/d^{2}{\bf q})$. The above results in
Eq. (\ref{effective-spin-excitation}) show that the effective spin propagator
$P({\bf k},{\bf p},{\bf k}',\omega)$ in Eq. (\ref{reduced-propagator}) scales with
$p^{2}$. Moreover, when the electron umklapp scattering kicks in, the temperature
scale that is proportional to $\Delta^{2}_{p}$ can be very low \cite{Ma23,Lee21}
due to the presence of this $p^{2}$ scaling in the effective spin propagator
(\ref{reduced-propagator}). In this case, $T_{\rm scale}=\bar{a}\Delta^{2}_{p}$
can be identified as the temperature scale, with the average value
$\bar{a}=(1/N)\sum\limits_{\bf q}a({\bf q})$ that is a constant at a given electron
doping. In particular, our numerical result indicates that the temperature scale
$T_{\rm scale}=\bar{a}\Delta^{2}_{p}=0.01247J\approx 12$K at the electron doping
concentration $\delta=0.19$, also in good agreement with the crossover temperature
shown in Fig. \ref{resistivity-temperature}.

According to the above temperature scale $T_{\rm scale}$, we now follows the
analyses carried out in the hole-doped case \cite{Ma23} to show that the
temperature region in the electron-doped side shown in
Fig. \ref{resistivity-temperature} can also be divided into three characteristic
regions: \\
(i) the low-temperature region, where $T> T_{\rm scale}+\omega_{{\bf k}_{\rm A}}$
with the spin excitation energy $\omega_{{\bf k}_{\rm A}}\approx 0.0022J\sim 2$K at
the $[\pi,\pi]$ point of BZ. In this low-temperature region, the kernel function
$F(\theta,\theta')$ can be reduced as $F(\theta,\theta')\propto T$, which generates
a low-temperature T-linear resistivity $\rho(T)\propto T$ as shown in
Fig. \ref{resistivity-temperature}; \\
(ii) the far-lower-temperature region, where $T< T_{\rm scale}$. In this
far-lower-temperature region, the kernel function $F(\theta,\theta')$ can be reduced
as $F(\theta,\theta')\propto T^{2}$, which leads to a T-quadratic resistivity
$\rho(T)\propto T^{2}$ as shown in Fig. \ref{resistivity-temperature};\\
(iii) the crossover temperature region, where
$T_{\rm scale}<T<T_{\rm scale}+\omega_{{\bf k}_{\rm A}}$. In this
extremely-narrow-crossover temperature region, the resistivity does not exhibit a
T-linear or a T-quadratic behavior, but instead shows a nonlinear behavior with
respect to temperature. The above three characteristic regions can be summarized
schematically in Fig. \ref{three-region}, where
$T^{*}_{\rm s}=T_{\rm scale}+\omega_{{\bf k}_{\rm A}}$, and
$T_{\rm s}=T_{\rm scale}$.

We now turn to show why the strength of the low-temperature T-linear resistivity
correlates with the corresponding magnitude of $T_{\rm c}$ in the overdoped
electron-doped cuprate superconductors to form the scaling relation shown
Fig. \ref{A1-Tc}: (i) in the framework of the kinetic-energy-driven
superconductivity \cite{Feng15,Feng0306,Feng12,Feng15a}, the main ingredient is
identified into an electron pairing mechanism involving {\it the spin excitation},
where both the electron pair gap (then the anomalous self-energy in the
particle-particle channel) and the single-particle coherence (then the normal
self-energy in the particle-hole channel) arise from the interaction between
electrons mediated by {\it the spin excitation}. In this case, the SC-state is
controlled by both the electron pair gap and the single-particle coherence, which
leads to that the maximal $T_{\rm c}$ occurs around the optimal electron doping,
and then decreases in both the underdoped and the overdoped regimes as shown in
Fig. \ref{Tc-doping}; (ii) on the other hand, the results presented here show that
the low-temperature T-linear resistivity in the overdoped regime shown
in Fig. \ref{resistivity-temperature} arises from the momentum relaxation due to
the electron umklapp scattering mediated by {\it the same spin excitation}, and
then the strength of the low-temperature T-linear resistivity
decreases as $T_{\rm c}$ decreases as shown in Fig. \ref{coefficient-doping}. In
other words, this {\it same spin excitation} that mediates both the pairing
electrons responsible for superconductivity and the electron umklapp
scattering responsible for the low-temperature T-linear resistivity
induces a correlation between the strength of the low-temperature T-linear
resistivity and the corresponding magnitude of $T_{\rm c}$ in the
overdoped electron-doped cuprate superconductors as shown in Fig. \ref{A1-Tc}.

Superconductivity in the electron-doped cuprate superconductors with the highest
$T_{\rm c}$ emerges directly as an instability of the strange metallic normal-state
\cite{Horio16,Song17,Lin20,Horio20b}, where the electronic state exhibits the
non-Fermi-liquid behaviour. It was realized that the fundamental principles must be
involved in the understanding of the overdoped strange-metal regime
\cite{Keimer15,Varma20,Phillips22}. However, if we look more closely at cuprate
superconductors, then there is a lot of evidences in favour of the $t$-$J$ model as
the basic underlying microscopic model \cite{Rice97}. More specifically, the ARPES
experimental observations \cite{Kim98,Campuzano04,Damascelli03,Fink07} indicate
that the $t$-$J$ model (\ref{tjham})
is of particular relevance to the low energy features of cuprate superconductors.
In this case, the non-Fermi-liquid behaviour of cuprate superconductors have been
studied based directly on the $t$-$J$ model
\cite{Hybertsen90,Gooding94,Kim98,Lee06,Brinckmann01,Spalek22,Corboz14,Sorella02,Edegger07,Edegger06,White98,Nagaosa90,Lee88},
where it has been shown that the non-Fermi-liquid behaviour is due to strong
electron scattering mediated by the spin fluctuations. Starting from the $t$-$J$
model (\ref{tjham}), we \cite{Tan21,Tan20,Mou17} have also studied recently the
intrinsic features of the electronic structure of the electron-doped cuprate
superconductors, where the coupling of a spin excitation to electron quasiparticles
leads to the emergence of the peak-dip-hump structure in the quasiparticle
excitation spectrum and the kink in the quasiparticle dispersion, in agreement with
the corresponding experimental data \cite{Horio16,Song17,Lin20,Horio20b}. These
results of the low-energy electronic structure \cite{Tan21,Tan20,Mou17} together
with the present results of the low-temperature T-linear resistivity in the
overdoped regime show that the same spin excitation that is responsible for pairing
the electrons also dominantly scatters the electrons in the overdoped strange-metal
phase responsible for the low-energy electronic structure. All these results
\cite{Tan21,Hybertsen90,Gooding94,Kim98,Lee06,Tan20,Mou17,Spalek22,Corboz14,Sorella02,Edegger07,Edegger06,White98,Nagaosa90,Lee88}
therefore indicates that the strong scattering between electrons mediated by the
spin excitations plays a crucial role in the understanding of the non-Fermi-liquid
behaviour in the overdoped electron-doped cuprate superconductors.

\section{Summary}\label{summary}

Within the framework of the kinetic-energy-driven superconductivity, we have
rederived the doping dependence of $T_{\rm c}$ in the electron-doped cuprate
superconductors, where the glue to hold the constrained electron pairs together is
{\it the spin excitation, the collective mode from the internal spin degree of
freedom of the constrained electron itself}, then $T_{\rm c}$ achieves its maximum
at around the optimal electron doping, and decreases in both the underdoped and
overdoped regimes. By virtue of this doping dependence of $T_{\rm c}$, we then have
investigated the correlation between the strength of the low-temperature T-linear
resistivity and the corresponding magnitude of $T_{\rm c}$ in the overdoped
electron-doped cuprate superconductors, where the low-temperature T-linear
resistivity in the overdoped regime arises from the momentum relaxation due to the
electron umklapp scattering mediated by {\it the same spin excitation}. The result
of the low-temperature T-linear resistivity in this paper together with the
previous work for superconductivity \cite{Tan21} therefore identify explicitly a
fact: the {\it spin excitation} that acts like a bosonic glue to hold the electron
pairs together responsible for superconductivity also mediates the electron
umklapp scattering responsible for the low-temperature T-linear resistivity in the
overdoped regime. This {\it same spin excitation} therefore leads to that the
strength of the low-temperature T-linear resistivity well tracks the corresponding
magnitude of $T_{\rm c}$ in the overdoped electron-doped cuprate superconductors.

After intensive investigations about four decades, although the underlying
scattering mechanism for the low-temperature resistivity of cuprate superconductors
still remains controversial, the electron umklapp scattering is believed to be at
the heart of the exotic features of the low-temperature resistivity in cuprate
superconductors \cite{Ma23,Lee21,Rice17,Hussey03}. In particular, it has been shown
that the low-temperature T-linear resistivity in the overdoped strange-metal phase
originates from the electron umklapp scattering \cite{Lee21}. On the other hand, it
has been shown that the electron umklapp scattering processes, which directly
transfer momentum between the electron sea and the underlying square lattice, lead
to the T-linear resistivity in the strange-metal phase \cite{Rice17}. More
specifically, the momentum relaxation due to the electron umklapp scattering
mediated by the spin excitations has been employed recently to study the nature of
the low-temperature T-linear resistivity in the overdoped strange-metal phase
\cite{Ma23}, where the underlying scattering rate is T-linear near the umklapp
point, which therefore leads to a low-temperature T-linear resistivity. All the
results obtained from these studies \cite{Ma23,Lee21,Rice17,Hussey03} are well
consistent with the corresponding experimental observations, and therefore confirm
that the low-temperature T-linear resistivity in the strange-metal phase
\cite{Greene20,Fournier98,Jin11,Sarkar17,Sarkar19,Legros19,Yuan22,Daou09,Cooper09,Ayres21,Grisso21}
arises from the electron umklapp scattering.

Finally, it should be noted that the effective spin propagator
$P({\bf k},{\bf p},{\bf k}',\omega)$ in Eq. (\ref{ESP}) is obtained in the
mean-field (MF) level \cite{Cheng08}, i.e., $P({\bf k},{\bf p},{\bf k}',\omega)$ in
Eq. (\ref{ESP}) is obtained in terms of the convolution of two MF spin propagators
in Eq. (\ref{MF-spin-propagator}). In this case, the umklapp scattering between
electrons in Eq. (\ref{electron-collision-1}) by the exchange of the effective MF
spin propagator is better suited for the interpretation of the low-temperature
T-linear resistivity in the strange-metal phase of the overdoped electron-doped
cuprate superconductors, as it has been done for the case in the overdoped
hole-doped cuprate superconductors \cite{Ma23}. This follows a basic fact that as
the results shown in Refs. \onlinecite{Feng15} and \onlinecite{Feng12}, the coupling strength
of the electrons with the spin excitations gradually weakens with the increase of
doping from a strong-coupling case in the underdoped regime to a weak-coupling side
in the overdoped regime \cite{Taillefer10,Johnson01,Dahm09,Kordyuk10}, reflecting
a reduction of the strength of the AF fluctuation with the increase of doping. In
other words, the effect of the AF fluctuation in the overdoped regime is less
dramatic than in the underdoped regime
\cite{Taillefer10,Wilson06,Wilson06-1,Wilson06-2}, leading to that the AF (then the
normal-state) pseudogap effect created by the AF fluctuation is quite weak
\cite{Horio16,Xu23a,Xu24}. However, in the underdoped regime, the effect of the AF
fluctuation is much dramatic \cite{Taillefer10,Wilson06,Wilson06-1,Wilson06-2}, and
can be described in terms of the full spin propagator \cite{Cheng08}, where the
spin self-energy is derived in terms of the charge-carrier bubble. In particular,
the AF (then the normal-state) pseudogap effect generated by the dramatic AF
fluctuation in the underdoped regime is quite strong \cite{Horio16,Xu23a,Xu24}, and
then this pseudogap lowers the density of the spin excitations in response to the
electron umklapp scattering. In this case, the electron umklapp scattering should
be mediated by the exchange of the effective full spin propagator for a proper
description of the low-temperature resistivity in the underdoped electron-doped
cuprate superconductors. These and the related issues are under investigation now.
In particular, the low-temperature T-quadratic resistivity of the underdoped
hole-doped cuprate superconductors has been discussed very recently \cite{Ma24},
where the scattering rate arises from the electron umklapp scattering by the exchange
of the effective full spin propagator. The obtained results of the low-temperature
T-quadratic resistivity in the underdoped regime \cite{Ma24} together with the study
on the low-temperature T-linear resistivity in the overdoped regime \cite{Ma23} show
that the the electron umklapp scattering from a spin excitation responsible for the
low-temperature T-linear resistivity in the overdoped regime naturally produces the
low-temperature T-quadratic resistivity in the underdoped regime resulting from the
opening of a momentum dependent spin pseudogap. This underlying mechanism of the
electron umklapp scattering from a spin excitation \cite{Ma23,Ma24} is also confirmed
from the experimental analyses \cite{Puchkov96,Bucher93,Ito93,Nakano94,Ando01}, where
it has been shown that if the electron scattering responsible for the low-temperature
T-linear resistivity in the overdoped regime involves the electron scattering on the
spin excitations in the underdoped regime, then the spin pseudogap seen in nuclear
magnetic resonance and nuclear quadrupole resonance below the normal-state pseudogap
crossover temperature would naturally account for a deviation from the low-temperature
T-linear behaviour of the resistivity. Since the common scattering mechanism linking
the low-temperature resistivity of both the hole- and electron-doped cuprate
superconductors, the present theory also predicts that the variation of the
low-temperature resistivity in the underdoped electron-doped cuprate superconductors
deviates from the T-linear behaviour, while this deviation from the low-temperature
T-linear behaviour of the resistivity has been observed experimentally in the
underdoped electron-doped cuprate superconductors
\cite{Poniatowski21a,Greene20,Fournier98,Jin11,Sarkar17,Sarkar19}.

\section*{Acknowledgements}

X. Ma would like to thank Faculty of Arts and Sciences, Beijing Normal University
at Zhuhai for the hospitality.

\section*{Disclosure statement}

No potential conflict of interest was reported by the authors.

\section*{Funding}

XM, MZ, and SP are supported by the National Key Research and Development Program of
China under Grant Nos. 2023YFA1406500 and 2021YFA1401803, and the National Natural
Science Foundation of China under Grant Nos. 12274036 and 12247116. HG acknowledges
support from the National Natural Science Foundation of China under Grant
Nos. 11774019 and 12074022.


\begin{appendix}

\section{Derivation of superconducting transition temperature $T_{\rm c}$}
\label{Derivation-of-Tc}

In this Appendix, our main goal is to generalize the theoretical framework of the
kinetic-energy-driven superconductivity from the previous hole-doped
case \cite{Feng9404,Feng15,Feng0306,Feng12,Feng15a} to the present electron-doped
case, and then derive the full hole diagonal and off-diagonal propagators
$G_{\rm f}({\bf k},\omega)$ and $\Im^{\dagger}_{\rm f}({\bf k},\omega)$ in
Eq. (\ref{HDODGF-2}) of the main text.

\subsection{Mean-field theory}\label{Mean-field-theory}

In the MF level, the $t$-$J$ model (\ref{cssham}) in the main text can be
decoupled as,
\begin{subequations}\label{MF-t-J-model}
\begin{eqnarray}
H_{\rm MF}&=&H_{\rm t}+H_{\rm J}+H_{0}, \\
H_{\rm t}&=&-\chi_{1}t\sum_{<ll'>\sigma}a^{\dagger}_{l'\sigma}a_{l\sigma}
+\chi_{2}t'\sum_{<<ll'>>\sigma}a^{\dagger}_{l'\sigma}a_{l\sigma}\nonumber\\
&+&\mu_{\rm a}\sum_{l\sigma}a^{\dagger}_{l\sigma}a_{l\sigma}, ~~~~~~~~
\label{MF-t-term}\\
H_{\rm J}&=& {1\over 2}J_{\rm eff}\sum_{<ll'>}[\epsilon(S^{+}_{l}S^{-}_{l'}
+S^{-}_{l}S^{+}_{l'})+2S^{\rm z}_{l}S^{\rm z}_{l'}]\nonumber\\
&+& t'\phi_{2}\sum_{<<ll'>>}(S^{+}_{l}S^{-}_{l'}+S^{-}_{l}S^{+}_{l'}),~~~~~~~~~
\label{MF-J-term}
\end{eqnarray}
\end{subequations}
where $H_{0}=8Nt\phi_{1}\chi_{1}-8Nt'\phi_{2}\chi_{2}$,
$\epsilon=1-2t\phi_{1}/J_{\rm eff}$, the charge-carrier's particle-hole parameters
$\phi_{1}=\langle a^{\dagger}_{l\sigma}a_{l+\hat{\eta}\sigma}\rangle$ and
$\phi_{2}=\langle a^{\dagger}_{l\sigma}a_{l+\hat{\tau}\sigma}\rangle$, and the
spin correlation functions $\chi_{1}=\langle S_{l}^{+}S_{l+\hat{\eta}}^{-}\rangle$
and $\chi_{2}=\langle S_{l}^{+}S_{l+\hat{\tau}}^{-}\rangle$, with
$\hat{\eta}=\hat{x},~\hat{y}$ and $\hat{\tau}=\hat{x}\pm\hat{y}$.

According to the above charge-carrier part (\ref{MF-t-term}), the MF charge-carrier
propagator can be obtained straightforwardly as,
\begin{eqnarray}\label{MFHGF}
g^{(0)}_{\rm a}({\bf k},\omega)={1\over \omega-\xi^{(\rm a)}_{\bf k}},
\end{eqnarray}
with the MF charge-carrier band energy,
\begin{eqnarray}\label{MFCCS}
\xi^{(\rm a)}_{\bf k}=-4\chi_{1}t\gamma_{\bf k}+4t'\chi_{2}\gamma_{\bf k}'
+\mu_{\rm a}.
\end{eqnarray}

On the other hand, in the doped regime without AFLRO, i.e.,
$\langle S^{\rm z}_{l}\rangle =0$, the spin propagator can be evaluated within
the Kondo-Yamaji decoupling scheme \cite{Kondo72}, which is a stage one-step
further than the Tyablikov's decoupling scheme \cite{Tyablikov67}. In particular,
in the MF level, the spin part in Eq. (\ref{MF-J-term}) is described by an
anisotropic Heisenberg model. In this case, two spin propagators $D(l-l',t-t')
=-i\theta(t-t')\langle [S^{+}_{l}(t),S^{-}_{l'}(t')]\rangle=\langle\langle
S^{+}_{l}(t);S^{-}_{l'}(t')\rangle\rangle$ and $D_{\rm z}(l-l',t-t')=-i\theta(t-t')
\langle [S^{\rm z}_{l}(t),S^{\rm z}_{l'}(t')]\rangle=\langle\langle
S^{\rm z}_{l}(t);S^{\rm z}_{l'}(t')\rangle\rangle$ are needed to give a proper
description of the nature of the spin excitation \cite{Cheng08}. Following these
previous discussions \cite{Cheng08}, the spin propagators $D^{(0)}({\bf k},\omega)$
and $D^{(0)}_{\rm z}({\bf k},\omega)$ can be respectively obtained as,
\begin{subequations}\label{TWO-MFSGF}
\begin{eqnarray}
D^{(0)}({\bf k},\omega)&=&{B_{\bf k}\over 2\omega_{\bf k}}\left ({1\over\omega-
\omega_{\bf k}}-{1\over\omega+\omega_{\bf k}}\right ),\label{MFSGF}\\
D^{(0)}_{\rm z}({\bf k},\omega)&=& {B_{{\rm z}{\bf k}}\over
2\omega_{{\rm z}{\bf k}}}\left ({1\over\omega-\omega_{{\rm z}{\bf k}}}
-{1\over\omega+\omega_{{\rm z}{\bf k}}}\right ),~~~~
\label{MFSGFZ}
\end{eqnarray}
\end{subequations}
as quoted in Eq. (\ref{MF-spin-propagator}) of the main text, with the weight
functions,
\begin{subequations}
\begin{eqnarray}
B_{\bf k}&=&\lambda_{1}[2\chi^{\rm z}_{1}(\epsilon\gamma_{\bf k}-1)
+\chi_{1}(\gamma_{\bf k}-\epsilon)]\nonumber\\
&-&\lambda_{2}(2\chi^{\rm z}_{2}
\gamma_{\bf k}'-\chi_{2}), ~~~~~\\
B_{{\rm z}{\bf k}} &=&\epsilon\chi_{1}\lambda_{1}(\gamma_{\bf k}-1)-\chi_{2}
\lambda_{2}(\gamma_{\bf k}'-1),
\end{eqnarray}
\end{subequations}
respectively, while the spin excitation spectra $\omega_{\bf k}$ and
$\omega_{{\rm z}{\bf k}}$ that are given by,
\begin{widetext}
\begin{subequations}\label{TWO-MFSES}
\begin{eqnarray}
\omega^{2}_{\bf k}&=&\lambda_{1}^{2}\left [{1\over 2}\epsilon\left (A_{1}
-{1\over 2}\alpha\chi^{\rm z}_{1}-\alpha\chi_{1}\gamma_{\bf k}\right)(\epsilon
-\gamma_{\bf k})+\left (A_{2}-{1\over 2Z}\alpha\epsilon\chi_{1}-\alpha\epsilon
\chi^{\rm z}_{1}\gamma_{\bf k}\right )
(1-\epsilon\gamma_{\bf k})\right ]\nonumber\\
&+&\lambda_{2}^{2}\left [\alpha
\left (\chi^{\rm z}_{2}\gamma_{\bf k}'-{3\over 2Z}\chi_{2}\right )\gamma_{\bf k}'
+{1\over 2}\left (A_{3}-{1\over 2}\alpha\chi^{\rm z}_{2}\right )\right ]
+\lambda_{1}\lambda_{2}\left [\alpha\chi^{\rm z}_{1}(1-\epsilon\gamma_{\bf k})
\gamma_{\bf k}'+{1\over 2}\alpha(\chi_{1}\gamma_{\bf k}'-C_{3})(\epsilon
-\gamma_{\bf k})\right .\nonumber\\
&+& \left . \alpha \gamma_{\bf k}'(C^{\rm z}_{3}-\epsilon\chi^{\rm z}_{2}
\gamma_{\bf k})
- {1\over 2}\alpha\epsilon (C_{3}-\chi_{2}\gamma_{\bf k})\right ],
\label{MFSES}\\
\omega^{2}_{{\rm z}{\bf k}}&=&\epsilon\lambda^{2}_{1}\left (\epsilon A_{1}
-{1\over Z}\alpha\chi_{1}-\alpha\chi_{1}\gamma_{\bf k}\right )(1-\gamma_{\bf k})
+\lambda^{2}_{2}A_{3}(1-\gamma_{\bf k}')
+\lambda_{1}\lambda_{2}[\alpha\epsilon C_{3}(\gamma_{\bf k}-1)+\alpha(\chi_{2}
\gamma_{\bf k}-\epsilon C_{3})(1 -\gamma_{\bf k}')],~~~~~\label{MFSESZ}
\end{eqnarray}
\end{subequations}
respectively, where $\lambda_{1}=2ZJ_{\rm eff}$, $\lambda_{2}=4Z\phi_{2}t'$,
$A_{1}=\alpha C_{1}+(1-\alpha)/(2Z)$, $A_{2}=\alpha C^{\rm z}_{1}+(1-\alpha)/(4Z)$,
$A_{3}=\alpha C_{2}+(1-\alpha) /(2Z)$, the spin correlation functions
$\chi^{\rm z}_{1}=\langle S_{l}^{\rm z}S_{l+\hat{\eta}}^{\rm z}\rangle$,
$\chi^{\rm z}_{2}=\langle S_{l}^{\rm z}S_{l+\hat{\tau}}^{\rm z}\rangle$,
$C_{1}=(1/Z^{2})\sum_{\hat{\eta},\hat{\eta'}}\langle S_{l+\hat{\eta}}^{+}
S_{l+\hat{\eta'}}^{-}\rangle$, $C^{\rm z}_{1}=(1/Z^{2})\sum_{\hat{\eta},\hat{\eta'}}
\langle S_{l+\hat{\eta}}^{z}S_{l+\hat{\eta'}}^{z}\rangle$,
$C_{2}=(1/Z^{2})\sum_{\hat{\tau},\hat{\tau'}}\langle S_{l+\hat{\tau}}^{+}
S_{l+\hat{\tau'}}^{-}\rangle$, $C_{3}=(1/Z)\sum_{\hat{\tau}}\langle
S_{l+\hat{\eta}}^{+}S_{l+\hat{\tau}}^{-}\rangle$, $C^{\rm z}_{3}=(1/Z)
\sum_{\hat{\tau}}\langle S_{l+\hat{\eta}}^{\rm z}S_{l+\hat{\tau}}^{\rm z}\rangle$,
and $Z=4$ is the number of the NN or next NN sites. In order to fulfill the sum rule
of the correlation function $\langle S^{+}_{l}S^{-}_{l}\rangle=1/2$ in the case
without an AFLRO, the important decoupling parameter $\alpha$ has been introduced in
the above calculation \cite{Cheng08,Kondo72}, which can be regarded as the vertex
correction.
\end{widetext}

\subsection{Charge-carrier diagonal and off-diagonal propagators}
\label{Charge-carrier-self-energy}

\begin{figure}[h!]
\centering
\includegraphics[scale=0.65]{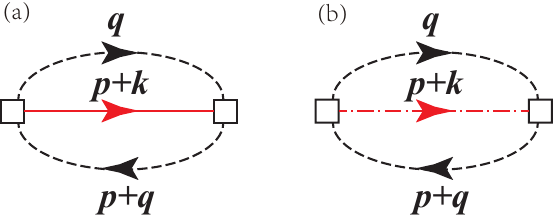}
\caption{The skeletal diagrams for the charge-carrier (a) normal and (b) anomalous
self-energies for scattering charge carriers from the spin excitations. The
red-solid-line and red-dash-dot-line represent the charge-carrier diagonal and
off-diagonal propagators $g_{\rm a}$ and $\Gamma^{\dagger}_{\rm a}$, respectively,
and the black-dash-line depicts the spin propagator $D^{(0)}$, while $\square$
describes the vertex function $\Lambda$. \label{charge-carrier-self-energy-diagram}}
\end{figure}

In the hole-doped case \cite{Feng9404,Feng15,Feng0306,Feng12}, it has been
demonstrated that the interaction between the charge carriers directly from the
kinetic energy of the $t$-$J$ model by the exchange of a spin excitation induces
the charge-carrier pairing state in the particle-particle channel. With the help
of these discussions \cite{Feng9404,Feng15,Feng0306,Feng12}, the full charge-carrier
diagonal and off-diagonal propagators of the $t$-$J$ model (\ref{cssham}) in the
charge-carrier pairing state satisfy the self-consistent Dyson's equations, which
can be derived in terms of the Eliashberg's approach \cite{Eliashberg60} as,
\begin{subequations}\label{CCSCES}
\begin{eqnarray}
g_{\rm a}({\bf k},\omega)&=&g^{(0)}_{\rm a}({\bf k},\omega)
+g^{(0)}_{\rm a}({\bf k},\omega)[\Sigma^{(\rm a)}_{\rm ph}({\bf k},\omega)
g_{\rm a}({\bf k},\omega)\nonumber\\
&-& \Sigma^{(\rm a)}_{\rm pp}({\bf k},\omega)
\Gamma^{\dagger}_{\rm a}({\bf k},\omega)],~~~~~~~~\\
\Gamma^{\dagger}_{\rm a}({\bf k},\omega)&=& g^{(0)}_{\rm a}({\bf k},-\omega)
[\Sigma^{(\rm a)}_{\rm ph}({\bf k},-\omega)\Gamma^{\dagger}_{\rm a}({\bf k},\omega)
\nonumber\\
&+&\Sigma^{(\rm a)}_{\rm pp}({\bf k},\omega)g_{\rm a}({\bf k},\omega)],~~~~
\end{eqnarray}
\end{subequations}
in the present electron-doped case, and then the full charge-carrier diagonal and
off-diagonal propagators $g_{\rm a}({\bf k},\omega)$ and
$\Gamma^{\dagger}_{\rm a}({\bf k},\omega)$ can be obtained directly as,
\begin{subequations}\label{CCDODGF}
\begin{eqnarray}
g_{\rm a}({\bf k},\omega)&=&{1\over\omega-\xi^{(\rm a)}_{\bf k}
-\Sigma^{(\rm a)}_{\rm tot}({\bf k},\omega)},~~~~~\label{CCDGF-1}\\
\Gamma^{\dagger}_{\rm a}({\bf k},\omega)&=&{L^{(\rm a)}({\bf k},\omega)\over\omega
-\xi^{(\rm a)}_{\bf k}-\Sigma^{(\rm a)}_{\rm tot}({\bf k},\omega)},
~~~~~~~\label{CCODGF-1}
\end{eqnarray}
\end{subequations}
respectively, where the charge-carrier total self-energy
$\Sigma^{(\rm a)}_{\rm tot}({\bf k},\omega)$ and the function
$L^{(\rm a)}({\bf k},\omega)$ can be expressed as,
\begin{subequations}\label{CCTOTSE}
\begin{eqnarray}
\Sigma^{(\rm a)}_{\rm tot}({\bf k},\omega)&=&
\Sigma^{(\rm a)}_{\rm ph}({\bf k},\omega)
+{|\Sigma^{(\rm a)}_{\rm pp}({\bf k},\omega)|^{2}\over\omega+\xi^{(\rm a)}_{\bf k}
+\Sigma^{(\rm a)}_{\rm ph}({\bf k},-\omega)},~~~~~\\
L^{(\rm a)}({\bf k},\omega)&=&-{\Sigma^{(\rm a)}_{\rm pp}({\bf k},\omega)\over
\omega+\xi^{(\rm a)}_{\bf k}+\Sigma^{(\rm a)}_{\rm ph}({\bf k},-\omega)},
\end{eqnarray}
\end{subequations}
respectively,
with the charge-carrier normal self-energy
$\Sigma^{(\rm a)}_{\rm ph}({\bf k},\omega)$ in the particle-hole channel sketched
in Fig. \ref{charge-carrier-self-energy-diagram}a and the charge-carrier anomalous
self-energy $\Sigma^{(\rm a)}_{\rm pp}({\bf k},\omega)$ in the particle-particle
channel sketched in Fig. \ref{charge-carrier-self-energy-diagram}b that have been
derived in terms of the spin bubble as \cite{Cheng08},
\begin{subequations}\label{CCSE}
\begin{eqnarray}
\Sigma^{({\rm a})}_{\rm ph}({\bf k},i\omega_{n})&=&{1\over N}\sum_{\bf p}
{1\over \beta}\sum_{ip_{m}}g_{\rm a}({\bf p}+{\bf k},ip_{m}+i\omega_{n})\nonumber\\
&\times& P^{(0)}({\bf k},{\bf p},ip_{m}), ~~~~~~~~~\\
\Sigma^{({\rm a})}_{\rm pp}({\bf k},i\omega_{n})&=&{1\over N}\sum_{\bf p}
{1\over \beta}\sum_{ip_{m}}
\Gamma^{\dagger}_{\rm a}({\bf p}+{\bf k},ip_{m}+i\omega_{n})\nonumber\\
&\times& P^{(0)}({\bf k},{\bf p},ip_{m}),
\end{eqnarray}
\end{subequations}
respectively,
where the effective spin propagator,
\begin{eqnarray}\label{ESP-A}
P^{(0)}({\bf k},{\bf p},\omega)={1\over N}\sum_{\bf q}
\Lambda^{2}_{{\bf p}+{\bf q}+{\bf k}}\Pi({\bf p},{\bf q},\omega),
\end{eqnarray}
as quoted in Eq. (\ref{ESP-1}) of the main text, is closely related to the spin
bubble $\Pi({\bf p},{\bf q},\omega)$ in Eq. (\ref{spin-bubble}), and has been
evaluated in subsection \ref{Effective-propagator} of the main text.

The above results in Eqs. (\ref{CCSCES}) and (\ref{CCSE}) indicate that the
charge-carrier normal and anomalous self-energies
$\Sigma^{({\rm a})}_{\rm ph}({\bf k},\omega)$ and
$\Sigma^{({\rm a})}_{\rm pp}({\bf k},\omega)$ self-consistently link up with the
full charge-carrier diagonal and off-diagonal propagators
$g_{\rm a}({\bf k},\omega)$ and $\Gamma^{\dagger}_{\rm a}({\bf k},\omega)$. In
particular, $\Sigma^{(\rm a)}_{\rm ph}({\bf k},\omega)$ is defined as the momentum
and energy dependence of the charge-carrier pair gap, i.e.,
$\Sigma^{({\rm a})}_{\rm pp}({\bf k},\omega)=\bar{\Delta}^{\rm (a)}_{\bf k}(\omega)$,
while $\Sigma^{(\rm a)}_{\rm ph}({\bf k},\omega)$ depicts the momentum and energy
dependence of the charge-carrier quasiparticle coherence, which therefore competes
with charge-carrier pairing-state.

$\Sigma^{({\rm a})}_{\rm pp}({\bf k},\omega)$ is an even function of energy,
however, $\Sigma^{({\rm a})}_{\rm ph}({\bf k},\omega)$ is not. In this case,
$\Sigma^{({\rm a})}_{\rm ph}({\bf k},\omega)$ can be separated into its symmetric
and antisymmetric parts as: $\Sigma^{({\rm a})}_{\rm ph}({\bf k},\omega)
=\Sigma^{({\rm a})}_{\rm phe}({\bf k},\omega)
+\omega\Sigma^{({\rm a})}_{\rm pho}({\bf k},\omega)$, and then both
$\Sigma^{({\rm a})}_{\rm phe}({\bf k},\omega)$ and
$\Sigma^{({\rm a})}_{\rm pho}({\bf k},\omega)$ are an even function of energy.
In particular, this antisymmetric part
$\Sigma^{({\rm a})}_{\rm pho}({\bf k},\omega)$ is associated closely with the
momentum and energy dependence of the charge-carrier quasiparticle coherent weight
as: $Z^{{\rm (a)}-1}_{\rm F}({\bf k},\omega)
=1-{\rm Re}\Sigma^{(\rm a)}_{\rm pho}({\bf k},\omega)$. In this paper, we focus
mainly on the low-energy behaviors, and then
$\bar{\Delta}^{\rm (a)}_{\bf k}(\omega)$ and $Z^{{\rm (a)}}_{\rm F}({\bf k},\omega)$
can be generally discussed in the static limit as,
\begin{subequations}
\begin{eqnarray}
\bar{\Delta}^{\rm (a)}_{\bf k}&=&\bar{\Delta}^{\rm (a)}_{\bf k}(\omega=0)
=\bar{\Delta}_{\rm a}\gamma^{\rm (d)}_{\bf k}, \label{CCPGF}\\
{1\over Z^{\rm (a)}_{\rm F}({\bf k})}&=&1-
{\rm Re}\Sigma^{(\rm a)}_{\rm pho}({\bf k},\omega=0),\label{CCQCWW}
\end{eqnarray}
\end{subequations}
respectively, with the charge-carrier pair gap parameter $\bar{\Delta}_{\rm a}$
and the d-wave factor
$\gamma^{\rm (d)}_{\bf k}=({\rm cos}k_{x}-{\rm cos} k_{y})/2$.
Although $Z^{\rm (a)}_{\rm F}({\bf k})$ still is a function of momentum, the
momentum dependence is unimportant in a qualitative discussion. Following the
ARPES experiments \cite{DLFeng00,Ding01}, the momentum
${\bf k}$ in $Z^{\rm (a)}_{\rm F}({\bf k})$ can be chosen as,
\begin{eqnarray}\label{CCQCW}
Z^{\rm (a)}_{\rm F}=Z^{\rm (a)}_{\rm F}({\bf k})\mid_{{\bf k}=[\pi,0]}.
\end{eqnarray}
With the help of the above static-limit approximation, the renormalized
charge-carrier diagonal and off-diagonal propagators can be derived from
Eq. (\ref{CCDODGF}) as,
\begin{subequations}\label{BCSHGF}
\begin{eqnarray}
g^{\rm (RMF)}_{\rm a}({\bf k},\omega)&=&Z^{\rm (a)}_{\rm F}
\left ( {U^{2}_{{\rm a}{\bf k}}\over\omega-E^{\rm (a)}_{\bf k}}
+{V^{2}_{{\rm a}{\bf k}}\over\omega+E^{\rm (a)}_{\bf k}}\right ),
\label{BCSHDGF}\\
\Gamma_{\rm a}^{{\rm (RMF)}\dagger}({\bf k},\omega)&=&-Z^{\rm (a)}_{\rm F}
{\bar{\Delta}^{\rm (a)}_{{\rm Z}{\bf k}}\over 2E^{\rm (a)}_{\bf k}}
\left ( {1\over \omega-E^{\rm (a)}_{\bf k}}
-{1\over\omega+E^{\rm (a)}_{\bf k}}\right ),\nonumber\\
~~~~~~~~~~\label{BCSHODGF}
\end{eqnarray}
\end{subequations}
where $E^{\rm (a)}_{\bf k}=\sqrt{\bar{\xi}^{(\rm a)2}_{\bf k}
+\mid\bar{\Delta}^{\rm (a)}_{{\rm Z}{\bf k}}\mid^{2}}$ is the charge-carrier
quasiparticle energy dispersion,
$\bar{\xi}^{(\rm a)}_{{\bf k}}=Z^{\rm (a)}_{\rm F}\xi^{(\rm a)}_{\bf k}$ is the
renormalized MF charge-carrier band energy,
$\bar{\Delta}^{\rm (a)}_{{\rm Z}{\bf k}}=Z^{\rm (a)}_{\rm F}
\bar{\Delta}^{\rm (a)}_{\bf k}$ is the renormalized charge-carrier pair gap, and
the charge-carrier quasiparticle coherence factors $U_{{\rm a}{\bf k}}$ and
$V_{{\rm a}{\bf k}}$ are given by,
\begin{subequations}\label{BCSCF}
\begin{eqnarray}
U^{2}_{{\rm a}{\bf k}}&=&{1\over 2}\left (1+{\bar{\xi}^{(\rm a)}_{{\bf k}}\over
E^{\rm (a)}_{\bf k}}\right ),
\end{eqnarray}
\begin{eqnarray}
V^{2}_{{\rm a}{\bf k}}&=&{1\over 2}\left (1-{\bar{\xi}^{(\rm a)}_{{\bf k}}\over
E^{\rm (a)}_{\bf k}}\right ),
\end{eqnarray}
\end{subequations}
respectively. The above these charge-carrier quasiparticle coherence factors satisfy
the constraint $U^{2}_{{\rm a}{\bf k}}+V^{2}_{{\rm h}{\bf k}}=1$ for any momentum
${\bf k}$.

We now substitute the above charge-carrier renormalized diagonal and off-diagonal
propagators in Eq. (\ref{BCSHGF}) and spin propagator in Eq. (\ref{MFSGF}) into
Eq. (\ref{CCSE}), and then obtain explicitly the charge-carrier normal self-energy
$\Sigma^{({\rm a})}_{\rm ph}({\bf k},\omega)$ and the anomalous self-energy
$\Sigma^{({\rm a})}_{\rm pp}({\bf k},\omega)$ as,
\begin{widetext}
\begin{subequations}\label{SE1}
\begin{eqnarray}
{\Sigma}^{\rm(a)}_{\rm ph}({\bf{k}},\omega)&=&{Z^{\rm (a)}_{\rm F}\over N^{2}}
\sum_{{\bf{p}}{\bf{p}'}{\nu}}(-1)^{\nu+1}\Omega_{{\bf{p}}{\bf{p}'}{\bf{k}}}\left [
U^{2}_{\rm{a}{\bf p}+{\bf k}}\left ( {F^{(\rm{a})}_{1\nu}({\bf p},{\bf p}',{\bf k})
\over\omega+\omega^{(\nu)}_{{\bf p}{\bf p}'}-E^{\rm (a)}_{{\bf p}+{\bf k}}}
- {F^{(\rm{a})}_{2\nu}({\bf p},{\bf p}',{\bf k})\over\omega
-\omega^{(\nu)}_{{\bf p}{\bf p}'}-E^{\rm (a)}_{{\bf p}+{\bf k}}}\right )\right. \nonumber\\
&+& \left . V^{2}_{{\rm a}{\bf p}+{\bf k}}\left (
{F^{(\rm{a})}_{1\nu}({\bf p},{\bf p}',{\bf k})\over\omega
-\omega^{(\nu)}_{{\bf p}{\bf p}'}+E^{\rm (a)}_{{\bf p}+{\bf k}}}
-{F^{(\rm{a})}_{2\nu}({\bf p},{\bf p}',{\bf k})\over\omega
+\omega^{(\nu)}_{{\bf p}{\bf p}'}+E^{\rm (a)}_{{\bf p}+{\bf k}}}\right ) \right ],
~~~~ \\
{\Sigma}^{\rm(a)}_{\rm pp}({\bf{k}},\omega)&=&{Z^{\rm (a)}_{\rm F}\over N^{2}}
\sum_{{\bf{p}}{\bf{p}'}{\nu}}(-1)^{\nu}\Omega_{{\bf{p}}{\bf{p}'}{\bf{k}}}
{\bar{\Delta}^{\rm (a)}_{{\rm Z}{\bf p}+{\bf k}}\over 2E^{\rm (a)}_{{\bf p}+{\bf k}}}
\left [ \left (
{F^{(\rm{a})}_{1\nu}({\bf p},{\bf p}',{\bf k})\over\omega
+\omega^{(\nu)}_{{\bf p}{\bf p}'}-E^{\rm (a)}_{{\bf p}+{\bf k}}}
- {F^{(\rm{a})}_{2\nu}({\bf p},{\bf p}',{\bf k})\over\omega
-\omega^{(\nu)}_{{\bf p}{\bf p}'}-E^{\rm (a)}_{{\bf p}+{\bf k}}}\right )\right .\nonumber\\
&-& \left. \left ( {F^{(\rm{a})}_{1\nu}({\bf p},{\bf p}',{\bf k})\over\omega
-\omega^{(\nu)}_{{\bf p}{\bf p}'}+E^{\rm (a)}_{{\bf p}+{\bf k}}}
- {F^{(\rm{a})}_{2\nu}({\bf p},{\bf p}',{\bf k})\over\omega
+\omega^{(\nu)}_{{\bf p}{\bf p}'}+E^{\rm (a)}_{{\bf p}+{\bf k}}}\right ) \right ],
\end{eqnarray}
\end{subequations}
respectively, with $\nu=1,~2$, ${\Omega}_{{\bf{p}}{\bf{p}'}{\bf{k}}}
=\Lambda^{2}_{{\bf{p}}+{\bf{p}'}+{\bf{k}}}B_{\bf{p}'}
B_{\bf{p}+\bf{p}'}/(4\omega_{{\bf p}'}\omega_{{\bf p}+{\bf p}'})$,
$\omega^{(\nu)}_{{\bf p}{\bf p}'}={\omega}_{\bf{p}+\bf{p}'}
-(-1)^{\nu}{\omega}_{\bf{p}'}$, and the functions,
\begin{subequations}
\begin{eqnarray}
F^{(\rm{a})}_{1\nu}({\bf p},{\bf p}',{\bf k})&=&n_{\rm{F}}
[E^{\rm (a)}_{{\bf p}+{\bf k}}]\{1+n_{\rm B}(\omega_{\bf{p}'+\bf{p}})
+ n_{\rm B}[(-1)^{\nu+1}\omega_{\bf{p}'}]\}
+n_{\rm B}(\omega_{\bf{p}'+\bf{p}})n_{\rm B}[(-1)^{\nu+1}\omega_{\bf{p}'}],\\
F^{(\rm{a})}_{2\nu}({\bf{p}},{\bf{p}'},{\bf{k}})&=&\{1-n_{\rm{F}}
[E^{\rm (a)}_{{\bf{p}}+{\bf{k}}}]\}\{1+n_{\rm B}(\omega_{\bf{p}'+\bf{p}})
+ n_{\rm B}[(-1)^{\nu+1}\omega_{\bf{p}'}]\}
+n_{B}(\omega_{\bf{p}'+\bf{p}})n_{B}[(-1)^{\nu+1}\omega_{\bf{p}'}].~~~~~~~~
\end{eqnarray}
\end{subequations}

\subsection{Self-consistent equations for determination of charge-carrier pair gap
parameter and the related order parameters}\label{CSCE}

The above charge-carrier pair gap parameter $\bar{\Delta}_{\rm a}$ and the
charge-carrier quasiparticle coherent weight $Z^{\rm (a)}_{\rm F}$ satisfy following
two self-consistent equations,
\begin{subequations}\label{SCE1}
\begin{eqnarray}
\bar{\Delta}_{a}&=& {4Z^{\rm (a)2}_{\rm F}\over N^{3}}
\sum_{{\bf{p}}{\bf{p}'}{\bf{k}}{\nu}}(-1)^{\nu}
\Omega_{{\bf{p}}{\bf{p}'}{\bf{k}}}{\gamma^{\rm (d)}_{\bf k}\bar{\Delta}_{\rm a}
\gamma^{\rm (d)}_{{\bf p}+{\bf k}}\over E^{\rm (a)}_{\bf{p}+\bf{k}}}
\left( {F^{(\rm{a})}_{1\nu}({\bf{p}},{\bf{p}}',{\bf{k}})\over
\omega^{(\nu)}_{{\bf{p}}{\bf{p}}'}-E^{\rm (a)}_{\bf{p}+\bf{k}}}
- {F^{(\rm{a})}_{2\nu}({\bf{p}},{\bf{p}}',{\bf{k}})\over
\omega^{(\nu)}_{{\bf{p}}{\bf{p}}'}+E^{\rm (a)}_{\bf{p}+\bf{k}}} \right ),
\nonumber\\
~~~~~~\\
{1\over Z^{\rm (a)}_{\rm F}} &=& 1+{Z^{\rm (a)}_{\rm F}\over N^{2}}
\sum_{{\bf{p}}{\bf{p}'}{\nu}}(-1)^{\nu+1}\Omega_{{\bf{p}}{\bf{p}'}{{\bf k}_{\rm A}}}
\left ( {F^{(\rm{a})}_{1\nu}({\bf{p}},{\bf{p}}',{\bf k}_{\rm A})\over
[\omega^{(\nu)}_{{\bf{p}}{\bf{p}}'}-E^{\rm (a)}_{\bf{p}+{\bf k}_{\rm A}}]^{2}}
+{F^{(\rm{a})}_{2\nu}({\bf{p}},{\bf{p}}',{\bf k}_{\rm A})\over
[\omega^{(\nu)}_{{\bf{p}}{\bf{p}}'}+E^{\rm (a)}_{\bf{p}+\bf{k}_{\rm A}}]^{2}}
\right ), \nonumber\\
\end{eqnarray}
\end{subequations}
respectively, where ${\bf k}_{\rm A}=[\pi,0]$. These two equations must be solved
simultaneously with following self-consistent equations,
\end{widetext}
\begin{subequations}\label{SCE2}
\begin{eqnarray}
\phi_{1}&=&{Z^{\rm (a)}_{\rm F}\over 2N}\sum_{\bf k}\gamma_{\bf k}\left (
1-{\bar{\xi}^{(\rm a)}_{\bf{k}}\over E^{\rm (a)}_{\bf k}}{\rm{tanh}}\left [
{1\over 2}\beta E^{\rm (a)}_{\bf k}\right ] \right ), ~~~~~\\
\phi_{2}&=&{Z^{\rm (a)}_{\rm F}\over 2N}\sum_{{\bf{k}}}\gamma_{{\bf{k}}}'\left (
1-{\bar{\xi}^{(\rm a)}_{\bf{k}}\over E^{\rm (a)}_{\bf k}}{\rm{tanh}}\left [
{1\over 2}\beta E^{\rm (a)}_{\bf k}\right ] \right),~~~~~
\end{eqnarray}
\begin{eqnarray}
\delta &=&{Z^{\rm (a)}_{\rm F}\over 2N}\sum_{{\bf{k}}}\left (
1-{\bar{\xi}^{(\rm a)}_{\bf{k}}\over E^{\rm (a)}_{\bf k}}{\rm{tanh}}\left [
{1\over 2}\beta E^{\rm (a)}_{\bf k}\right ] \right ),~~~~~\\
\chi_{1}&=&{1\over N}\sum_{{\bf{k}}}\gamma_{\bf k}{B_{\bf{k}}\over
2\omega_{\bf{k}}}\rm{coth}\left [ {1\over 2}\beta\omega_{\bf{k}}\right ],
\end{eqnarray}
\begin{eqnarray}
\chi_{2}&=&{1\over N}\sum_{{\bf{k}}}\gamma_{{\bf{k}}}'{B_{\bf{k}}\over
2\omega_{\bf{k}}}\rm{coth}\left [ {1\over 2}\beta\omega_{\bf{k}}\right ],\\
C_{1}&=&{1\over N}\sum_{{\bf{k}}}\gamma^{2}_{\bf k}{B_{\bf{k}}\over
2\omega_{\bf{k}}}\rm{coth}\left [ {1\over 2}\beta\omega_{\bf{k}}\right ],\\
C_{2}&=&{1\over N}\sum_{{\bf{k}}}\gamma'^{2}_{\bf{k}}{B_{\bf{k}}\over
2\omega_{\bf{k}}}\rm{coth}\left [ {1\over 2}\beta\omega_{\bf{k}}\right],\\
C_{3}&=&{1\over N}\sum_{{\bf{k}}}\gamma_{\bf k}\gamma_{\bf{k}'}{B_{\bf{k}}\over
2\omega_{\bf{k}}}\rm{coth}\left [ {1\over 2}\beta\omega_{\bf{k}}\right],\\
{1\over 2}&=&{1\over N}\sum_{{\bf{k}}}{B_{\bf{k}}\over 2\omega_{\bf{k}}}\rm{coth}
\left [ {1\over 2}\beta\omega_{\bf{k}}\right],\\
\chi^{\rm z}_{1}&=&{1\over N}\sum_{{\bf{k}}}\gamma_{\bf k}{B_{{\rm z}{\bf k}}\over
2\omega_{{\rm z}{\bf k}}}\rm{coth}\left [ {1\over 2}\beta\omega_{{\rm z}{\bf k}}
\right],  \\
\chi^{\rm z}_{2}&=&{1\over N}\sum_{{\bf{k}}}\gamma_{{\bf{k}}'}{B_{{\rm z}{\bf k}}
\over 2\omega_{{\rm z}{\bf k}}}\rm{coth}\left [ {1\over 2}\beta
\omega_{{\rm z}{\bf k}}\right],\\
C^{\rm z}_{1}&=&{1\over N}\sum_{{\bf{k}}}\gamma^{2}_{\bf k}{B_{{\rm z}{\bf k}}\over
2\omega_{{\rm z}{\bf k}}}\rm{coth}\left [ {1\over 2}\beta\omega_{{\rm z}{\bf k}}
\right],  \\
C^{\rm z}_{3}&=&{1\over N}\sum_{{\bf{k}}}\gamma_{\bf k}\gamma_{\bf{k}'}
{B_{{\rm z}{\bf k}}\over 2\omega_{{\rm z}{\bf k}}}\rm{coth}\left [ {1\over 2}\beta
\omega_{{\rm z}{\bf k}} \right],
\end{eqnarray}
\end{subequations}
then the charge-carrier pair gap parameter and the related order parameters, the
decoupling parameter $\alpha$, and the charge-carrier chemical potential
$\mu_{\rm a}$ are determined by the self-consistent calculation without using any
adjustable parameters.

In the previous discussions \cite{Tan20,Tan21,Mou17}, the above equations
(\ref{SCE1}) and (A19) have been evaluated self-consistently, and the
obtained result of the evolution of the charge-carrier pair gap parameter
$\bar{\Delta}_{\rm a}$ in Eq. (\ref{CCPGF}) with doping indicates that with the
increase of doping, $\bar{\Delta}_{\rm a}$ is raised gradually in the underdoped
regime, and then reaches the maximum at around the optimal doping
$\delta\approx 0.15$. However, with the further increase of doping,
$\bar{\Delta}_{\rm a}$ then turns into a monotonically decrease in the overdoped
regime.

\subsection{Charge-carrier pair transition temperature}\label{CCPTT}

The charge-carrier pair transition temperature $T^{\rm (pair)}_{\rm c}$ as a
function of doping \cite{Tan21} has been calculated self-consistently from the
above self-consistent equations (\ref{SCE1}) and (A19) at the condition of
the charge-carrier pair gap parameter $\bar{\Delta}_{\rm a}=0$, and the obtained
result shows that $T^{\rm (pair)}_{\rm c}$ as a function of doping presents a
similar behavior of $\bar{\Delta}_{\rm a}$, i.e., the maximal
$T^{\rm (pair)}_{\rm c}$ occurs around the optimal doping $\delta\approx 0.15$,
and then decreases in both the underdoped and the overdoped regimes.

In the framework of the kinetic-energy driven superconductivity
\cite{Feng15,Feng0306,Feng12,Feng15a}, the spin excitation is directly coupled to
charge-carrier pairs, where the interaction between the charge carriers in the
particle-particle channel is attractive, then the system of charge carriers forms
pairs of bound charge carriers. Since the coupling strength $V_{\rm eff}$ and
charge-carrier pair order parameter $\Delta_{\rm a}$ have been incorporated into
the charge-carrier pair gap parameter $\bar{\Delta}_{\rm a}$, the coupling
strength $V_{\rm eff}$ can be evaluated in terms of the ratio of
$\bar{\Delta}_{\rm a}$ and $\Delta_{\rm a}$ as \cite{Feng15,Feng12},
\begin{eqnarray}\label{strength}
V_{\rm eff}={\bar{\Delta}_{\rm a}\over\Delta_{\rm a}},
\end{eqnarray}
where the charge-carrier pair order parameter $\Delta_{\rm a}$ can be derived
directly from the charge-carrier off-diagonal propagator (\ref{BCSHODGF}) as,
\begin{eqnarray}\label{CCDGP}
\Delta_{\rm a}={2\over N}\sum_{{\bf k}}[\gamma^{(\rm d)}_{{\bf k}}]^{2}
{[Z^{\rm (a)}_{\rm F}]^{2}\bar{\Delta}_{\rm a}\over E^{\rm (a)}_{\bf k}}{\rm tanh}
\left({1\over 2}\beta E^{\rm (a)}_{\bf k} \right).~~~~
\end{eqnarray}
The parent compounds of the electron-doped cuprate superconductors are a Mott
insulator \cite{Fujita12,Anderson87} as we have mentioned in above Section
\ref{Introduction}. When charge carriers are doped into this parent Mott insulator,
there is a gain in the kinetic energy per charge carrier proportional to $t$ due to
hopping, however, at the same time, the magnetic energy is decreased, costing an
energy of approximately $J$ per site \cite{Lee06}, therefore the doped charge
carriers into a Mott insulator can be considered as a competition between the
kinetic energy ($\delta t$) and magnetic energy ($J$). In this case, the essential
physics of the dome-like shape doping dependence of $T^{\rm (pair)}_{\rm c}$ can be
attributed to the competition between the kinetic energy and magnetic energy
\cite{Feng15,Feng12}. This follows a basic fact that the competition between the
kinetic energy and magnetic energy also leads to that the spectral intensity of the
spin excitation spectrum decreases with the increase of doping \cite{Cheng08}.
However, as we \cite{Feng15,Feng12} have shown in the previous discussions that a
decrease of the spin excitation spectral intensity with the increase of doping
also leads to a decrease of the coupling strength $V_{\rm eff}$ with the increase
of doping, i.e., the coupling strength $V_{\rm eff}$ smoothly decreases with the
increase of doping from a strong-coupling case in the underdoped regime to a
weak-coupling side in the overdoped regime \cite{Johnson01,Dahm09,Kordyuk10}. On
the other hand, the charge-carrier kinetic energy, which is proportional to doping
$\delta t$, gradually increases with the increase of doping. In the underdoped
regime, the coupling strength $V_{\rm eff}$ is very strong to bind the most charge
carriers into charge-carrier pairs, and thus the number of charge-carrier pairs
increases with the increase of doping, which leads to that
$T^{\rm (pair)}_{\rm c}$ increases with the increase of doping \cite{Feng15,Feng12}.
However, in the overdoped regime, the coupling
strength $V_{\rm eff}$ is relatively weak. In this case, not all charge carriers
can be bound to form charge-carrier pairs by this weakly attractive interaction,
and then the number of charge-carrier pairs decreases with the increase of doping,
which leads to that $T^{\rm (pair)}_{\rm c}$ decreases with the further increase of doping
\cite{Feng15,Feng12}. In particular, the optimal doping is a balance point, where
the number of charge-carrier pairs and coupling strength $V_{\rm eff}$ are
optimally matched \cite{Feng15,Feng12}. This is why $T^{\rm (pair)}_{\rm c}$ takes
a dome-like shape with the underdoped and overdoped regimes on each side of the
optimal doping, where $T^{\rm (pair)}_{\rm c}$ reaches its maximum \cite{Tan21}.

Furthermore, it has been shown that this $T^{\rm (pair)}_{\rm c}$ is the exactly
same as that derived from the corresponding SC-state at the condition of the SC gap
parameter $\bar{\Delta}=0$, and will return to this discussion of
$T^{\rm (pair)}_{\rm c}$ towards subsection \ref{EPTT} of this Appendix.

\subsection{Derivation of full hole diagonal and off-diagonal propagators}
\label{FCSRS}

In the previous studies for the hole-doped case \cite{Feng15a} with the on-site
local constraint $\sum_{\sigma}C^{\dagger}_{l\sigma}C_{l\sigma}\leq 1$, the full
charge-spin recombination scheme has been established, where a charge carrier and a
localized spin are fully recombined into a constrained electron. In this full
charge-spin recombination scheme \cite{Feng15a}, we have realized that the coupling
form between the electrons and spin excitations is the same as that between the
charge carriers and spin excitations, which therefore indicates that the form of the
self-consistent equations satisfied by the full electron diagonal and off-diagonal
propagators is the same as the form satisfied by the full charge-carrier diagonal
and off-diagonal propagators. In the present electron-doped case with the on-site
local constraint $\sum_{\sigma}f^{\dagger}_{l\sigma}f_{l\sigma}\leq 1$ in the hole
representation (\ref{tjham-hole-representation}), we \cite{Tan20,Tan21,Mou17}
follow these previous discussions for the hole-doped case \cite{Feng15a} and perform
a full charge-spin recombination in which the full charge-carrier diagonal and
off-diagonal propagators $g_{\rm a}({\bf k},\omega)$ and
$\Gamma^{\dagger}_{\varsigma}({\bf k},\omega)$ in
Eq. (\ref{CCSCES}) are replaced by the full hole diagonal and off-diagonal
propagators $G_{\rm f}({\bf k},\omega)$ and $\Im^{\dagger}_{\rm f}({\bf k},\omega)$,
respectively, and then the self-consistent Dyson's equations satisfied by the full
hole diagonal and off-diagonal propagators of the $t$-$J$ model
(\ref{tjham-hole-representation}) in the SC-state can be derived in terms of the
Eliashberg's approach \cite{Eliashberg60} as,
\begin{subequations}\label{ESCES}
\begin{eqnarray}
G_{\rm f}({\bf k},\omega)&=&G^{(0)}_{\rm f}({\bf k},\omega)
+G^{(0)}_{\rm f}({\bf k},\omega)[\Sigma^{(\rm f)}_{\rm ph}({\bf k},\omega)
G_{\rm f}({\bf k},\omega)\nonumber\\
&-&\Sigma^{(\rm f)}_{\rm pp}({\bf k},\omega)
\Im^{\dagger}_{\rm f}({\bf k},\omega)],\nonumber\\
~~~~~~~~~~\label{EDGF} \\
\Im^{\dagger}_{\rm f}({\bf k},\omega)&=&G^{(0)}_{\rm f}({\bf k},-\omega)
[\Sigma^{(\rm f)}_{\rm ph}({\bf k},-\omega)\Im^{\dagger}_{\rm f}({\bf k},\omega)
\nonumber\\
&+&\Sigma^{(\rm f)}_{\rm pp}({\bf k},\omega)G_{\rm f}({\bf k},\omega)], \label{EODGF}
\end{eqnarray}
\end{subequations}
where the hole non-interacting (diagonal) propagator of the $t$-$J$ model
$G^{(0)}_{\rm f}({\bf k},\omega)$ can be expressed as
$G^{(0)-1}_{\rm f}({\bf k},\omega)=\omega-\varepsilon^{\rm (f)}_{\bf k}$.
With the help of the above self-consistent equations (\ref{ESCES}), we can obtain
straightforwardly the full hole diagonal and off-diagonal propagators as,
\begin{subequations}\label{HDODGF-A}
\begin{eqnarray}
G_{\rm f}({\bf k},\omega)&=&{1\over\omega-\varepsilon^{({\rm f})}_{\bf k}
-\Sigma^{(\rm f)}_{\rm tot}({\bf k},\omega)},~~~~~\label{HDGF-A}\\
\Im^{\dagger}_{\rm f}({\bf k},\omega)&=&{L^{(\rm f)}({\bf k},\omega)\over\omega
-\varepsilon^{({\rm f})}_{\bf k}-\Sigma^{(\rm f)}_{\rm tot}({\bf k},\omega)},
~~~~~~~\label{HODGF-A}
\end{eqnarray}
\end{subequations}
as quoted in Eq. (\ref{HDODGF-2}) of the main text, where the total hole
self-energy $\Sigma^{(\rm f)}_{\rm tot}({\bf k},\omega)$ and the function
$L^{(\rm f)}({\bf k},\omega)$ have been given in Eq. \ref{HTOTSE} of the main text.
In particular, the hole normal
self-energy $\Sigma^{(\rm f)}_{\rm ph}({\bf k},\omega)$ in the particle-hole
channel sketched in Fig. \ref{hole-self-energy-diagram}a and hole anomalous
self-energy $\Sigma^{(\rm f)}_{\rm pp}({\bf k},\omega)$ in the particle-particle
channel sketched in Fig. \ref{hole-self-energy-diagram}b have been evaluated
directly from the corresponding parts of the charge-carrier normal self-energy
$\Sigma^{(\rm a)}_{\rm ph}({\bf k},\omega)$ and charge-carrier anomalous
self-energy $\Sigma^{(\rm a)}_{\rm pp}({\bf k},\omega)$ in Eq. (\ref{CCSE}) by the
replacement of the full charge-carrier diagonal and off-diagonal propagators
$g_{\rm a}({\bf k},\omega)$ and $\Gamma^{\dagger}_{\rm a}({\bf k},\omega)$ with the
corresponding full hole diagonal and off-diagonal propagators
$G_{\rm f}({\bf k},\omega)$ and $\Im^{\dagger}_{\rm f}({\bf k},\omega)$ as,
\begin{subequations}\label{ESE}
\begin{eqnarray}
\Sigma^{(\rm f)}_{\rm ph}({\bf k},i\omega_{n})&=&{1\over N}\sum_{\bf p}
{1\over \beta}\sum_{ip_{m}}G_{\rm f}({\bf p}+{\bf k},ip_{m}+i\omega_{n})\nonumber\\
&\times& P^{(0)}({\bf k},{\bf p},ip_{m}),
\label{HNSE-A}\\
\Sigma^{(\rm f)}_{\rm pp}({\bf k},i\omega_{n})&=&{1\over N}
\sum_{\bf p}{1\over \beta}\sum_{ip_{m}}
\Im^{\dagger}_{\rm f}({\bf p}+{\bf k},ip_{m}+i\omega_{n})\nonumber\\
&\times& P^{(0)}({\bf k},{\bf p},ip_{m}),\label{HASE-A}
\end{eqnarray}
\end{subequations}
respectively, as quoted in Eq. (\ref{HNASE-1}) of the main text.

In analogy to the derivation of the charge-carrier normal and anomalous
self-energies in subsection \ref{Charge-carrier-self-energy}, the hole normal and
anomalous self-energies have been also calculated explicitly in the previous
studies \cite{Tan21,Tan20,Mou17}. Following the discussions in subsection
\ref{Charge-carrier-self-energy}, the hole normal self-energy
$\Sigma^{(\rm f)}_{\rm ph}({\bf k},\omega)$ can be broken up into
its symmetric and antisymmetric parts as:
$\Sigma^{(\rm f)}_{\rm ph}({\bf k},\omega)=\Sigma^{(\rm f)}_{\rm phe}({\bf k},\omega)
+\omega\Sigma^{(\rm f)}_{\rm pho}({\bf k},\omega)$, where all the symmetric part
$\Sigma^{(\rm f)}_{\rm phe}({\bf k},\omega)$ and antisymmetric part
$\Sigma^{(\rm f)}_{\rm pho}({\bf k},\omega)$ are an even function of energy. Moreover,
the antisymmetric part $\Sigma^{(\rm f)}_{\rm pho}({\bf k},\omega)$ is identified as
the hole quasiparticle coherent weight: $Z^{(\rm f)-1}_{\rm F}({\bf k},\omega)
=1-{\rm Re}\Sigma^{(\rm f)}_{\rm pho}({\bf k},\omega)$. In an interacting system,
everything happens at around EFS. As a case for low-energy close to EFS, the hole pair
gap and hole quasiparticle coherent weight can be discussed in the static-limit
approximation,
\begin{subequations}
\begin{eqnarray}
\bar{\Delta}^{(\rm f)}_{\bf k}&=& \Sigma^{(\rm f)}_{\rm pp}({\bf k},\omega=0)=
\bar{\Delta}_{\rm f}\gamma^{(\rm d)}_{\bf k},\label{EPGF}\\
{1\over Z^{(\rm f)}_{\rm F}}&=&1-{\rm Re}\Sigma^{(\rm f)}_{\rm pho}({\bf k},\omega=0)
\mid_{{\bf k}=[\pi,0]}, ~~~\label{EQCW}
\end{eqnarray}
\end{subequations}
where $\bar{\Delta}_{\rm f}$ is the hole pair gap parameter, while the wave vector
${\bf k}$ in $Z^{(\rm f)}_{\rm F}({\bf k})$ has been chosen as ${\bf k}=[\pi,0]$ just
as it has been done in the ARPES experiments \cite{Ding01,DLFeng00}.

Based on the above static-limit approximation for the hole pair gap
$\bar{\Delta}^{(\rm f)}_{\bf k}$ and hole quasiparticle coherent weight
$Z^{(\rm f)}_{\rm F}$, we can obtain the renormalized hole diagonal and off-diagonal
propagators from Eq. (\ref{HDODGF-A}) as,
\begin{subequations}\label{MF-EGFS}
\begin{eqnarray}
G^{\rm (RMF)}_{\rm f}({\bf k},\omega)&=&Z^{(\rm f)}_{\rm F}\left (
{U^{2}_{{\rm f}{\bf k}}\over\omega-E^{(\rm f)}_{\bf k}}+{V^{2}_{{\rm f}{\bf k}}
\over\omega+E^{(\rm f)}_{\bf k}} \right ), \\
\Im^{{\rm (RMF)}\dagger}_{\rm f}({\bf k},\omega)&=&-Z^{(\rm f)}_{\rm F}
{\bar{\Delta}^{(\rm f)}_{\rm Z}({\bf k})\over 2E^{(\rm f)}_{\bf k}}\left (
{1\over\omega-E^{(\rm f)}_{\bf k}}-{1\over\omega+E^{(\rm f)}_{\bf k}}\right ),\nonumber\\
~~~
\end{eqnarray}
\end{subequations}
with the renormalized hole band energy $\bar{\varepsilon}^{(\rm f)}_{\bf k}
=Z^{(\rm f)}_{\rm F}\varepsilon^{(\rm f)}_{\bf k}$, the renormalized hole pair gap
$\bar{\Delta}^{(\rm f)}_{\rm Z}({\bf k})
=Z^{(\rm f)}_{\rm F}\bar{\Delta}^{(\rm f)}_{\bf k}$, the hole quasiparticle energy
spectrum $E^{(\rm f)}_{\bf k}=\sqrt {\bar{\varepsilon}^{(\rm f)2}_{\bf k}
+|\bar{\Delta}^{(\rm f)}_{\rm Z}({\bf k})|^{2}}$, and the hole quasiparticle
coherence factors
\begin{subequations}\label{EBCSCF}
\begin{eqnarray}
U^{2}_{{\rm f}{\bf k}}&=&{1\over 2}\left ( 1+{\bar{\varepsilon}^{(\rm f)}_{\bf k}
\over E^{(\rm f)}_{\bf k}}\right ), \\
V^{2}_{{\rm f}{\bf k}}&=&{1\over 2}\left ( 1-{\bar{\varepsilon}^{(\rm f)}_{\bf k}
\over E^{(\rm f)}_{\bf k}}\right ),
\end{eqnarray}
\end{subequations}
with the constraint $U^{2}_{{\rm f}{\bf k}}+V^{2}_{{\rm f}{\bf k}}=1$ for any wave
vector ${\bf k}$.

We now substitute the above renormalized hole diagonal and off-diagonal propagators
in Eq. (\ref{MF-EGFS}) and spin propagator in Eq. (\ref{MFSGF}) into
Eqs. (\ref{ESE}), and then obtain the hole normal and anomalous self-energies as,
\begin{widetext}
\begin{subequations}\label{ESE-1}
\begin{eqnarray}
{\Sigma}^{(\rm f)}_{\rm ph}({\bf{k}},{\omega})&=&\frac{Z^{(\rm f)}_{\rm{F}}}{N^{2}}
\sum_{{\bf{p}}{\bf{p}'}{\nu}}(-1)^{\nu+1}{\Omega}_{{\bf{p}}{\bf{p}'}{\bf{k}}}
\left [ U^{2}_{{\rm f}\bf{p}+\bf{k}}
\left ( \frac{F^{(\rm f)}_{1\nu}({\bf p},{\bf p}',{\bf k})}{\omega
+\omega^{(\nu)}_{{\bf{p}}{\bf{p}}'}-E^{(\rm f)}_{\bf{p}+\bf{k}}}
- \frac{F^{(\rm f)}_{2\nu}({\bf p},{\bf p}',{\bf k})}{\omega
-\omega^{(\nu)}_{{\bf{p}}{\bf{p}}'}-E^{(\rm f)}_{\bf{p}+\bf{k}}} \right)\right.
\nonumber\\
&+& \left . V^{2}_{{\rm f}\bf{p}+\bf{k}} \left (
\frac{F^{(\rm f)}_{1\nu}({\bf p},{\bf p}',{\bf k})}{\omega
-\omega^{(\nu)}_{{\bf{p}}{\bf{p}}'}+E^{(\rm f)}_{\bf{p}+\bf{k}}}
- \frac{F^{(\rm f)}_{2\nu}({\bf p},{\bf p}',{\bf k})}{\omega
+\omega^{(\nu)}_{{\bf{p}}{\bf{p}}'}+E^{(\varsigma)}_{\bf{p}+\bf{k}}}\right )
\right ], \label{ph-ESE}\\
{\Sigma}^{(\rm f)}_{\rm pp}({\bf{k}},{\omega})&=&\frac{Z^{(\rm f)}_{\rm{F}}}{N^{2}}
\sum_{{\bf{p}}{\bf{p}'}{\nu}}(-1)^{\nu}{\Omega}_{{\bf{p}}{\bf{p}'}{\bf{k}}}
\frac{\bar{\Delta}^{(\rm f)}_{\rm{Z}}({\bf{p}}+{\bf{k}})}
{2E^{(\rm f)}_{{\bf{p}}+\bf{k}}}\left [
\left ( \frac{F^{(\rm f)}_{1\nu}({\bf p},{\bf p}',{\bf k})}{\omega
+\omega^{(\nu)}_{{\bf{p}}{\bf{p}}'}-E^{(\rm f)}_{\bf{p}+\bf{k}}}
-\frac{F^{(\rm f)}_{2\nu}({\bf p},{\bf p}',{\bf k})}{\omega
-\omega^{(\nu)}_{{\bf{p}}{\bf{p}}'}-E^{(\rm f)}_{\bf{p}+\bf{k}}} \right)\right.
\nonumber\\
&-&\left. \left (
\frac{F^{(\rm f)}_{1\nu}({\bf p},{\bf p}',{\bf k})}{\omega
-\omega^{(\nu)}_{{\bf{p}}{\bf{p}}'}+E^{(\rm f)}_{\bf{p}+\bf{k}}}
- \frac{F^{(\rm f)}_{2\nu}({\bf p},{\bf p}',{\bf{k}})}{\omega
+\omega^{(\nu)}_{{\bf{p}}{\bf{p}}'}+E^{(\rm f)}_{\bf{p}+\bf{k}}}\right )\right ],
\label{pp-ESE}
\end{eqnarray}
\end{subequations}
respectively, with $\nu=1,~2$, and the functions,
\begin{subequations}
\begin{eqnarray}
F^{(\rm f)}_{1\nu}({\bf{p}},{\bf{p}}',{\bf{k}})&=&
n_{\rm{F}}(E^{(\rm f)}_{\bf{p}+\bf{k}})\{1+n_{B}(\omega_{\bf{p}'+\bf{p}})
+n_{B}[(-1)^{\nu+1}\omega_{\bf{p}'}]\}
+n_{B}(\omega_{\bf{p}'+\bf{p}})n_{B}[(-1)^{\nu+1}\omega_{\bf{p}'}], ~~~~\\
F^{(\rm f)}_{2\nu}({\bf{p}},{\bf{p}}',{\bf{k}})&=&
[1-n_{\rm{F}}(E^{(\rm f)}_{\bf{p}+\bf{k}})]\{1+n_{B}(\omega_{\bf{p}'+\bf{p}})
+n_{B}[(-1)^{\nu+1}\omega_{\bf{p}'}] \}
+n_{B}(\omega_{\bf{p}'+\bf{p}})n_{B}[(-1)^{\nu+1}\omega_{\bf{p}'}].~~~~
\end{eqnarray}
\end{subequations}

\subsection{Self-consistent equations for determination of hole order parameters}
\label{ESCE}

The above hole quasiparticle coherent weight, the hole pair gap parameter, and the
hole chemical potential fulfills following three self-consistent equations,
\begin{subequations}\label{SCE3}
\begin{eqnarray}
{1\over Z^{(\rm f)}_{\rm F}} &=& 1+{Z^{(\rm f)}_{\rm F}\over N^{2}}
\sum_{{\bf{p}}{\bf{p}'}{\nu}}(-1)^{\nu+1}\Omega_{{\bf{p}}{\bf{p}'}{\bf k}_{\rm A}}
\left (\frac{F^{(\rm f)}_{1\nu}({\bf p},{\bf p}',{\bf k}_{\rm A})}
{(\omega^{(\nu)}_{{\bf{p}}{\bf{p}}'}-E^{(\rm f)}_{\bf{p}+{\bf k}_{\rm A}})^{2}}
+\frac{F^{(\rm f)}_{2\nu}({\bf{p}},{\bf{p}}',{\bf k}_{\rm A})}
{(\omega^{(\nu)}_{{\bf{p}}{\bf{p}}'}+E^{(\rm f)}_{\bf{p}+{\bf k}_{\rm A}})^{2}}
\right ), \nonumber\\
~~~~~~~~~\\
\bar{\Delta}_{\rm f} &=& {4Z^{(\rm f)2}_{\rm F}\over N^{3}}
\sum_{{\bf{p}}{\bf{p}'}{\bf{k}}{\nu}}(-1)^{\nu}\Omega_{{\bf{p}}{\bf{p}'}{\bf{k}}}
\frac{\gamma^{\rm (d)}_{\bf k}\bar{\Delta}_{\rm f}
\gamma^{\rm (d)}_{{\bf p}+{\bf k}}}{E^{(\rm f)}_{\bf{p}+\bf{k}}}\left (
{F^{(\rm f)}_{1\nu}({\bf p},{\bf p}',{\bf k})\over\omega^{(\nu)}_{{\bf p}{\bf p}'}
-E^{(\rm f)}_{\bf{p}+\bf{k}}}-{F^{(\rm f)}_{2\nu}({\bf p},{\bf p}',{\bf k})\over
\omega^{(\nu)}_{{\bf p}{\bf p}'}+E^{(\rm f)}_{\bf{p}+\bf{k}}} \right ),\nonumber\\
~~~~~ \\
1-\delta &=&{Z^{(\rm f)}_{\rm{F}}\over N}\sum_{{\bf{k}}} \left(
1-{\bar{\varepsilon}^{(\rm f)}_{\bf k}\over E^{(\rm f)}_{\bf k}}\rm{tanh}
\left[ {1\over 2}\beta E^{(\rm f)}_{\bf k} \right ] \right ),\label{SCE-EFS}
~~~~~
\end{eqnarray}
\end{subequations}
\end{widetext}
where ${\bf k}_{\rm A}=[\pi,0]$. With the same calculation condition as in the
evaluation of the self-consistent equations (\ref{SCE1}) and (A19), the
above equations (\ref{SCE3}) have been also calculated self-consistently
\cite{Tan21,Tan20,Mou17}, where the hole quasiparticle coherent weight, the hole
pair gap parameter, and hole chemical potential are obtained without using any
adjustable parameters. In particular, the dome-like shape doping dependence of the
charge-carrier pair gap parameter $\bar{\Delta}_{\rm a}$ thus also induces the
same dome-like shape doping dependence of the hole pair gap parameter
$\bar{\Delta}_{\rm f}$.

\subsection{Superconducting transition temperature}\label{EPTT}

As we \cite{Tan21,Tan20,Mou17} have shown in Eq. (\ref{EASE}) of the main text,
the electron anomalous self-energy is obtained from the hole anomalous self-energy
in terms of the particle-hole transformation as:
$\Sigma_{\rm pp}({\bf k},\omega)=\Sigma^{({\rm f})}_{\rm pp}({\bf k},\omega)$,
which indicates that the electron pair gap $\bar{\Delta}_{\bf k}$ is identical with
the hole pair gap $\bar{\Delta}^{\rm (f)}_{\bf k}$,
i.e., $\bar{\Delta}_{\bf k}=\bar{\Delta}^{\rm (f)}_{\bf k}$. In this case, the SC
transition temperature $T_{\rm c}$ as a function of doping can be derived
self-consistently from the above self-consistent equations in Eq. (\ref{SCE3}) at
the condition of the hole pair gap $\bar{\Delta}^{\rm (f)}_{\bf k}=0$ [then the
electron pair gap $\bar{\Delta}_{\bf k}=0$]. In the recent discussions
\cite{Tan21}, we have found that in a given doping concentration, the magnitude
of the SC transition temperature $T_{\rm c}$ in Fig. \ref{Tc-doping} evaluated
from the above self-consistent equations in Eq. (\ref{SCE3}) is exactly identical
with the magnitude of the charge-carrier pair transition temperature
$T^{\rm (pair)}_{\rm c}$ calculated from the corresponding self-consistent
equations in Eqs. (\ref{SCE1}) and (A19) at the condition of the
charge-carrier pair gap $\bar{\Delta}^{\rm (a)}_{\bf k}=0$. This follows a basic
fact that in the framework of the kinetic-energy-driven superconductivity
\cite{Feng15,Feng0306,Feng12,Feng15a}, the effective attractive interaction
between charge carriers originates in their coupling to the spin excitation. On
the other hand, the hole (then the electron) pairing interaction in the full
charge-spin recombination scheme \cite{Feng15a} is mediated by the exchange of the
same spin excitation, which thus induces that the SC transition temperature
$T_{\rm c}$ shown in Fig. \ref{Tc-doping} is the same as the charge-carrier pair
transition temperature $T^{\rm (pair)}_{\rm c}$, and then the dome-like shape of
the doping dependence of $T_{\rm c}$ with its maximum occurring at around the
optimal doping is a natural consequence of the dome-like shape of the doping
dependence of $T^{\rm (pair)}_{\rm c}$ with its maximum occurring at around the
same optimal doping.

\end{appendix}

\end{document}